\documentclass{aastex63}
\usepackage[utf8]{inputenc}
\usepackage[T1]{fontenc}
\usepackage{comment}

\newcommand{\binary}{14 }
\newcommand{\ksdssobjs}{713 }
\newcommand{\ksdssrots}{179 }
\newcommand{\litrots}{7281 }

\newcommand{\prot}{\ensuremath{P_{\mbox{\scriptsize rot}}}}

\newcommand{\kms}{\ensuremath{\mbox{km s}^{-1}}}
\newcommand{\mas}{\ensuremath{\mbox{mas yr}^{-1}}}

\newcommand{\grp}{\ensuremath{(G - G_{\rm RP})}}

\shorttitle{Evaluating Rotation Periods of M dwarfs}
\shortauthors{Popinchalk et al.}

\begin{document}

\title{Evaluating Rotation Periods of M Dwarfs Across the Ages}

\correspondingauthor{Mark Popinchalk}
\author[0000-0001-9482-7794]{Mark Popinchalk}
\affiliation{Department of Astrophysics, American Museum of Natural History, Central Park West at 79th Street, New York, NY 10034, USA}
\affiliation{Physics, The Graduate Center, City University of New York, New York, NY 10016, USA}
\affiliation{Department of Physics and Astronomy, Hunter College, City University of New York, 695 Park Avenue, New York, NY 10065, USA}
\email{popinchalkmark@gmail.com}

\author[0000-0001-6251-0573]{Jacqueline K. Faherty}
\affiliation{Department of Astrophysics, American Museum of Natural History, Central Park West at 79th Street, New York, NY 10034, USA}

\author[0000-0003-2102-3159]{Rocio Kiman}
\affiliation{Department of Astrophysics, American Museum of Natural History, Central Park West at 79th Street, New York, NY 10034, USA}
\affiliation{Physics, The Graduate Center, City University of New York, New York, NY 10016, USA}
\affiliation{Department of Physics and Astronomy, Hunter College, City University of New York, 695 Park Avenue, New York, NY 10065, USA}

\author[0000-0002-2592-9612]{Jonathan Gagn\'e}
\affiliation{Plan\'etarium Rio Tinto Alcan, Espace pour la Vie, 4801 av. Pierre-de Coubertin, Montr\'eal, Qu\'ebec, Canada}
\affiliation{Institute for Research on Exoplanets, Universit\'e de Montr\'eal, D\'epartement de Physique, C.P.~6128 Succ. Centre-ville, Montr\'eal, QC H3C~3J7, Canada}

\author[0000-0002-2792-134X]{Jason L. Curtis}
\affiliation{Department of Astrophysics, American Museum of Natural History, Central Park West at 79th Street, New York, NY 10034, USA}
\affiliation{Department of Astronomy, Columbia University, 550 West 120th Street, New York, NY, USA}

\author[0000-0003-4540-5661]{Ruth Angus}
\affiliation{Department of Astrophysics, American Museum of Natural History, Central Park West at 79th Street, New York, NY 10034, USA}
\affiliation{Flatiron Institute, 162 Fifth Avenue, New York, NY 10010, USA}
\affiliation{Department of Astronomy, Columbia University, 550 West 120th Street, New York, NY, USA}

\author[0000-0002-1821-0650]{Kelle L. Cruz}
\affiliation{Department of Astrophysics, American Museum of Natural History, Central Park West at 79th Street, New York, NY 10034, USA}
\affiliation{Department of Physics and Astronomy, Hunter College, City University of New York, 695 Park Avenue, New York, NY 10065, USA}
\affiliation{Flatiron Institute, 162 Fifth Avenue, New York, NY 10010, USA}

\author[0000-0002-3252-5886]{Emily L. Rice}
\affiliation{Department of Astrophysics, American Museum of Natural History, Central Park West at 79th Street, New York, NY 10034, USA}
\affiliation{Macaulay Honors College, 35 W. 67th Street, New York, NY 10023, USA}

%%%%%%%%%%%%%%%%%%%%%%%%%%%%%%%%%%%%%%%%%%%%%%%%%%%%%%%%%%%%%%%%%%%%%%%%%%%%%%%%%%%%%%%%%%%%%%%%%%%%%%%%%%%%%%%%%%%%%%%%

\begin{abstract}
In this work we examine M dwarf rotation rates at a range of ages to establish benchmarks for M dwarf gyrochronology. This work includes a sample of \ksdssobjs spectroscopically-classified M0--M8 dwarfs with new rotation rates measured from K2 light curves. We analyzed data and recover rotation rates for \ksdssrots of these objects. 

We add these to rotation rates for members of clusters with known ages (5-700 Myr), as well as objects assumed to have field ages ($\gtrsim$1~Gyr). We use Gaia DR2 parallax and \grp\ photometry to create color--magnitude diagrams to compare objects across samples. We use color--period plots to analyze the period distributions across age, as well as incorporate H$\alpha$ equivalent width and tangential velocity where possible to further comment on age dependence. We find that the age of transition from rapid to slow rotation in clusters, which we define as an elbow in the period--color plots, depends on spectral type. Later spectral types transition at older ages: M4 for Praesepe at $\approx$700 Myr, one of the oldest clusters for which M dwarf rotation rates have been measured. The transition from active to inactive H$\alpha$ equivalent width also occurs at this elbow, as objects transition from rapid rotation to the slowly rotating sequence. Redder or smaller stars remain active at older ages. Finally, using Gaia kinematics we find evidence for rotation stalling for late Ms in the field sample, suggesting the transition happens much later for mid to late-type M dwarfs.
\end{abstract}
%%%%%%%%%%%%%%%%%%%%%%%%%%%%%%%%%%%%%%%%%%%%%%%%%%%%%%%%%%%%%%%%%%%%%%%%%%%%%%%%%%%%%%%%%%%%%%%%%%%%%%%%%%%%%%%%%%%%%%%%

\section{Introduction}

M dwarfs are the most numerous types of star in the galaxy as the peak of the initial mass function happens around M5 \citep{bochanski_2010,gould_1996}. They evolve on longer timescales than more massive stars, with their main sequence lifetimes extending beyond the current age of the universe \citep{laughlin_1997}. Several are known planet hosts, and M dwarfs have greater occurrence rates of rocky planets than more massive stars (e.g., \citet{hardegree_2019} and \citet{dressing_2015}). Their long lifetimes makes understanding the ages of M dwarfs an important parameter when studying them, or the exoplanets they host.

However, M dwarf ages are challenging to determine. Isochrone methods commonly used for more massive stars are not as effective due to their long stable main sequence lifetimes \citep{baraffe,soderblom_2010}. Within the M dwarf regime there is also a change in stellar interiors. While the majority of M dwarfs are above the hydrogen burning limit, M3-M4 provides an additional division between partially convective and fully convective interiors \citep{charb_baraffe}. This has consequences for the topology and strength of their magnetic fields \citep{charbon_dynamo}. Observations of M dwarf magnetic activity have shown a relationship with age as well as their rotation \citep{west_2015_activity_rotation,newton_2017_halpha}, and provide an opportunity for an empirical relation with age \citep[e.g.,]{skumanich_1972,pizzolato_2003_xraysat,barnes_2007_review,mamajek_hillenbrand}.

Additionally, the motions of older objects can be perturbed over time through interactions with molecular clouds passing stars and the galactic potential in general. This can lead to differences in velocities compared to younger objects, increasing velocity dispersion in populations, or even vertical action. This kinematic heating has proven to be a useful indicator of age in low mass objects.  \citet{faherty09} used tangential velocity ($v_{\rm tan}$) to discern older and younger populations of ultracool dwarfs. \citet{kiman_2019} found vertical action to be an age indicator for field M dwarfs. \citet{schmidt_2007} and \citet{gizis_2000} combined kinematics with magnetic activity for late M and L dwarfs. While overall they found a lack of correlation between magnetic activity strength and $v_{\rm tan}$ in these field M dwarfs, they did find evidence for the more active objects being younger. \citet{reid_1995} found that velocity dispersion increases with decreasing absolute magnitude in M dwarfs. Recently,  \citet{angus_2020} found an increase in velocity dispersion in the direction of galactic latitude for longer rotating stars for the entire Kepler sample \citep{mcq2014}.

One of the fundamental ways stars evolve in time is through changing rotation rates. \citet{skumanich_1972} identified a power law decrease in rotation rate with age in Sun-like stars. This was interpreted as angular momentum loss due to interactions between the magnetic field of a star and its stellar winds. \citet{barnes_2003} introduced the name gyrochronology, as well as the so-called \textit{I} sequence of slow rotators across color and \textit{C} sequence of fast rotators within clusters.  Ground-based surveys of clusters (eg. \citealt{irwin_2008_ngc2362}) have enabled age--rotation relationships across color (eg. \citealt{mamajek_hillenbrand,matt_2015,reiners_mohanty_2012}) often finding a Skumanich-like power law relation for the spindown of sun-like stars. However, these scaling relations are not complete, as \citet{angus2015} detailed: no single gyrochronology relation fits all clusters across spectral types, even for the better-behaved F,G,K stars. \citet{curtis2019} revealed that for the 1 Gyr old NGC 6811 cluster, there is a stalling that occurs in the G and K stars, as they lie on top of those in the 700 Myr old Praesepe cluster. This implies there could be additional issues for current gyrochronology models further along the low-mass end of stellar spectral types.

Specific attempts to constrain M dwarf gyrochronology include \citet{rebull2018usco}, who examined young cluster populations from $\sim$ 5-10 Myr through to approximately 700 Myr by comparing the slope of rotation periods across color between clusters of different ages. \citet{stauffer_bin_18} looked at the effects of binarity on rotation rate in low-mass stars within specific clusters. \citet{newton2016} used the MEarth  sample \citep{berta_mearth} to investigate the nearby field population and found evidence for a rapid transition in rotation rate from fast to slow for late M dwarfs, that was not apparent at the ages probed by \citet{rebull2018usco}. Combining these two perspectives proves difficult, as the ages involved are so different. Clearly, there are gaps in our knowledge for the rotational evolutionary timeline for M dwarfs.

The coupling of interior layers of stars has been proposed as a means to transfer angular momentum (example \citealt{dennissenkov}) which would imply that the change in interior structure to fully convective in  late-type Ms may be different to early types that have both radiative and convective layers. However, \citet{wright2018} found that fully convective M dwarfs are consistent with partially convective objects how their Rossby numbers relate to fractional X ray luminosities in the case of unsaturated stars, implying that the magnetic dynamos are still similar regardless of interior structure. Theory and observation have yet to come to an agreement on how the interior structure of these low-mass objects affects their rotational evolution.

Additional rotation rates are required to fully understand M dwarf gyrochronolgy. The intrinsic faintness in M dwarfs makes it difficult to gather high quality data for a large population of M dwarfs. Work has been done to identify M dwarfs in large survey missions such as SDSS (\citealt{west2011,schmidt_2015}), and the K2 mission \citep{howell_k2} provides months-long photometry for some objects. Furthermore, the second data release of the Gaia mission (Gaia DR2 hereafter) \citep{gaiadr2} is an invaluable resource for consistent photometry and parallaxes for a large number of M dwarfs in the solar neighborhood. 

The goal of this paper is to compile rotation rates for M dwarfs from the literature, while adding a sample of new measurements, to unify the available data and derive a consistent up-to-date picture of M dwarf gyrochronology. Previous work targeting the lowest mass stars have focused either on specific clusters \citep[e.g.,][]{rebull2018usco} or field samples \citep[e.g.,][]{newton2016,mcq_mdwarf}. We present a more holistic understanding of M dwarf rotation period distributions with time, by bringing both kinds of populations together in our analysis.

The sections of this paper are arranged as follows. In Section \ref{sec:sample}, we introduce the K2SDSS sample of M dwarfs defined from objects in both the K2 and SDSS missions. We also present a compilation of rotation rates for M dwarfs originating from the literature, and cross-match them with Gaia DR2. Section \ref{sec:rotation_k2sdss} presents our efforts to measure \ksdssrots new rotation rates for the K2SDSS sample, and includes comments on interesting light curves, such as candidate binaries. In Section \ref{5rotation_cmd}, we inspect all the objects with rotation rates in a Gaia color--magnitude diagram (CMD). In Section \ref{6rotation_age} we discuss M~dwarf rotation rate distributions across age. In section \ref{7halphakin}, we add additional context by including H$\alpha$ measurements and kinematics of the objects with rotation rates, where available. In Section \ref{sec:standardpic}, we discuss the standard M~dwarf gyrochronology picture in light of our analysis, before presenting conclusions in Section \ref{sec:conclusions}.
%%%%%%%%%%%%%%%%%%%%%%%%%%%%%%%%%%%%%%%%%%%%%%%%%%%%%%%%%%%%%%%%%%%%%%%%%%%%%%%%%%%%%%%%%%%%%%%%%%%%%%%%%%%%%%%%%%%%%%%%

\section{Sample}\label{sec:sample}

In this section we define a catalog of M dwarfs with measured rotation rates. 
We began by cross-matching the MLSDSS sample from \citet{kiman_2019} with K2. This allowed us to recover \ksdssobjs M dwarfs, which we use to define the K2SDSS sample (see Table \ref{k2sdssmemb_new}). We then compile a catalog of \litrots rotation rates for M dwarfs from the literature (see Table \ref{source}). The differences between these samples and considerations for comparing them are presented below.

\subsection{The K2SDSS subsample}\label{k2sdssintro}

In this section we introduce a new sample of stars for which we measured rotation rates, based on the MLSDSS sample from \citet{kiman_2019}. The MLSDSS sample is a curated sample of M dwarfs that have spectra from the Sloan Digital Sky Survey (SDSS); \citep{west2011,schmidt_2015} and are cross-matched with Gaia DR2 \citep{gaiadr2}. The sample was then further refined through modified quality cuts on the Gaia measurements described in Section 2.3 of \citet{kiman_2019}, and are designed to be more suitable for M dwarfs than the standard Gaia quality cuts. These include cuts on parallax error, removing poor astrometric solutions, and cuts on photometric excess and error. We selected stars that were assigned EPIC numbers during the K2 mission from the MLSDSS sample. We joined the Gaia Source ID column from MLSDSS with an existing 1" cross-match of Gaia DR2 with K2.\footnote{\url{gaia-kepler.fun}, see acknowledgments} This resulted in a sample of \ksdssobjs objects, which we designate as the K2SDSS sample. It includes objects from 14 K2 campaigns, with a K2 magnitude range of 13.75 to 23.24, and SDSS spectral types from M0 to L2. It is essential to this work that every object is associated with a Gaia Source ID with astrometric measurements, including parallax, and Gaia photometric bands \citep{gaiadr2}. Table \ref{k2sdssmemb_new} details the match between K2 and SDSS along with useful features compiled in the MLSDSS sample along with the Gaia measurements for each object.

\begin{deluxetable}{hcccchchccchchcccch}
\tabletypesize{\scriptsize}
\rotate
\tablecaption{K2SDSS Sample. \label{k2sdssmemb_new}}
\tablewidth{0pt}
\tablehead{
\nocolhead{SDSS Name} & \colhead{EPIC Number} & \colhead{Gaia DR2 Source ID}& \colhead{SPT} & \colhead{EWH$\alpha$}& \nocolhead{EW Error}& \colhead{RV} & \nocolhead{RV Error}& \colhead{Gaia R.A.} & \colhead{Gaia Decl.}& \colhead{pmRA} & \nocolhead{pmRA Error} &  \colhead{pmDec} & \nocolhead{pmDec Error} &  \colhead{$G$} & \colhead{$G_{\rm BP}$} & \colhead{$G_{\rm RP}$} & \colhead{Parallax} & \nocolhead{Parallax Error}\\
 & & & & (\AA) & (\AA)& (km/s) & (km/s)& (deg) & (deg) & (mas yr$^{-1}$) & (mas yr$^{-1}$) & (mas yr$^{-1}$) & (mas yr$^{-1}$) & (mag) & (mag) & (mag) & (mas) & (mas)}
\startdata
J044148.25+253430.2 &    247968420 & 148172179824515968 &  M7 & 235.661530 &  2.193080 &  19.169527 &       NaN &  70.451073 & 25.575074 &    4.511 &     0.400 &  $-$19.605 &      0.251 &      18.280 &       19.805 &       16.817 &  7.344 &    0.202 \\
J044110.79+255511.2 &    248015397 & 148196510814073728 &  M9 & 106.193253 &  0.468271 &  22.065651 & 10.136519 &  70.294965 & 25.919786 &    4.520 &     0.503 &  $-$20.112 &      0.294 &      18.138 &       20.744 &       16.573 &  6.377 &    0.254 \\
J043903.96+254426.0 &    247991214 & 148354733113981696 &  M9 &  60.881557 &  0.299452 &  27.444094 & 10.136519 &  69.766535 & 25.740561 &    7.037 &     0.362 &  $-$20.606 &      0.249 &      17.270 &       20.523 &       15.658 &  6.945 &    0.211 \\
J043947.48+260140.3 &    248029954 & 148400156688543104 &  M9 & 213.800858 &  0.843224 &  16.687094 &       NaN &  69.947874 & 26.027876 &    6.628 &     0.483 &  $-$21.865 &      0.309 &      17.780 &       20.880 &       16.009 &  6.797 &    0.237 \\
J085538.09+095258.4 &    211301854 & 597723696572386944 &  M8 &   3.375396 &  0.238983 & $-$29.995478 & 12.025241 & 133.908718 &  9.882913 &   $-$0.552 &     1.016 &  $-$16.805 &      0.707 &      19.647 &       21.673 &       18.167 &  6.905 &    0.653 \\
J085529.88+101713.7 &    211321707 & 597775167460827904 &  M9 &   6.131580 &  0.282073 & 110.190178 & 10.428387 & 133.874532 & 10.287166 &    9.692 &     0.677 &    4.200 &      0.436 &      19.167 &       21.759 &       17.581 & 14.803 &    0.432 \\
J083934.27+095931.2 &    211306999 & 598265785869780352 &  M7 &  $-$0.720708 &  0.257790 &  $-$6.688638 & 12.083542 & 129.892824 &  9.992021 &   26.772 &     0.626 &  $-$98.495 &      0.421 &      19.174 &       21.086 &       17.693 &  7.828 &    0.352 \\
J084027.86+100417.7 &    211310950 & 598270050772379008 &  M3 &  $-$0.157036 &  0.056579 &   9.622570 &  3.102581 & 130.116106 & 10.071602 &   19.681 &     0.144 &  $-$22.450 &      0.079 &      16.424 &       17.799 &       15.271 &  5.351 &    0.083 \\
J084333.38+102438.1 &    211328277 & 598390790892369920 &  L1 &   3.939189 &  1.492829 & $-$15.671266 & 12.533514 & 130.889087 & 10.410605 &  140.263 &     1.016 & $-$579.111 &      0.623 &      19.786 &       21.562 &       18.083 & 32.504 &    0.565 \\
J084414.70+105056.1 &    211351916 & 598789290843522560 &  M7 &  $-$0.344964 &  0.209632 &   2.827191 & 10.469080 & 131.061265 & 10.848922 &  $-$30.376 &     0.581 &  $-$22.219 &      0.406 &      19.017 &       20.729 &       17.559 &  8.676 &    0.343 \\
J084947.64+113353.8 &    211395034 & 598904503341393408 &  M0 &  $-$0.269399 &  0.040505 &   8.597193 &  3.898461 & 132.448531 & 11.564964 &  $-$10.730 &     0.123 &  $-$19.830 &      0.078 &      15.442 &       16.322 &       14.510 &  2.445 &    0.076 \\
J084938.64+113427.5 &    211395609 & 598905499773808768 &  M0 &  $-$0.379526 &  0.046850 & $-$22.037630 &  4.441122 & 132.411002 & 11.574326 &   $-$9.096 &     0.098 &  $-$15.856 &      0.065 &      15.993 &       16.815 &       15.107 &  1.949 &    0.059 \\
J084938.16+113535.1 &    211396891 & 598905602853022976 &  M0 &  $-$0.443856 &  0.064547 &  91.807022 &  6.074290 & 132.409028 & 11.593107 &  $-$12.742 &     0.169 &   $-$7.148 &      0.100 &      17.053 &       17.892 &       16.149 &  1.135 &    0.100 \\
J083155.92+102539.1 &    211329075 & 600076960693274496 &  M9 &   1.991869 &  0.249752 &  28.892153 & 10.858972 & 127.983001 & 10.427528 &  $-$64.856 &     0.580 & $-$171.808 &      0.444 &      18.176 &       20.827 &       16.549 & 31.563 &    0.296 \\
J082719.09+105444.2 &    211355470 & 600909600234425472 &  M3 &   0.499290 &  0.081143 &  48.380116 &  3.933673 & 126.829575 & 10.912282 &   47.747 &     0.229 &  $-$59.470 &      0.142 &      17.241 &       18.428 &       16.150 &  2.935 &    0.144 \\
J082344.57+103448.0 &    211337213 & 600976189407617920 &  M8 &   7.028934 &  0.208167 &  38.201088 & 10.152368 & 125.935743 & 10.580017 & $-$118.789 &     0.275 & $-$252.127 &      0.173 &      17.629 &       19.940 &       16.057 & 26.366 &    0.160 \\
J082542.62+111716.1 &    211377929 & 601122218295516032 &  M4 &  $-$0.044110 &  0.073050 &  58.400146 &  3.534853 & 126.427610 & 11.287809 &  $-$48.345 &     0.239 &  $-$55.212 &      0.131 &      17.158 &       18.700 &       15.923 &  3.986 &    0.150 \\
J082805.63+113359.6 &    211395130 & 601154409074776064 &  M8 &  11.954882 &  0.299089 &   8.067767 & 10.284334 & 127.023465 & 11.566582 &    6.809 &     1.110 &  $-$22.316 &      0.707 &      19.797 &       21.179 &       18.158 & 10.054 &    0.658 \\
J084224.39+104845.7 &    211349910 & 601421869573453952 &  M4 &   2.122004 &  0.145458 &   6.561838 &  2.339605 & 130.601643 & 10.812704 & $-$107.972 &     0.152 &   41.586 &      0.094 &      16.661 &       18.160 &       15.451 & 10.898 &    0.081 \\
J083916.75+105111.5 &    211352136 & 601456912210899968 &  M7 &   6.865015 &  0.319811 &  35.511845 & 11.602642 & 129.819803 & 10.853209 &  $-$24.135 &     0.906 &   $-$9.015 &      0.438 &      19.395 &       20.842 &       17.936 &  6.009 &    0.506 \\
\enddata
\tablecomments{Objects in the K2SDSS sample, drawn from a cross match between the K2 EPIC and known M dwarfs from SDSS. Spectral types (SPT), radial velocities (RV) and H$\alpha$ equivalent widths (EW H$\alpha$) come from SDSS spectra calculated by \citet{west2011,schmidt_2015}. Gaia DR2 values come from a cross match with SDSS. See \citet{kiman_2019} for further details. This is an example of the \ksdssobjs row table in the full online version.}
\end{deluxetable}

\subsection{A catalog of M dwarfs with rotation rates measured}\label{catalogM}

In this section we describe M dwarf rotation rates from the literature, and compile them in a single catalog. We introduce the studies that targeted specific clusters in Section \ref{source_clusters}, and field objects in Section \ref{source_field}, and describe our steps to match the objects with Gaia DR2 sources within each Section. Table \ref{source} summarizes the key features of all the literature samples included in this work. For each source, we list the target, the observatory or mission that produced the light curve, and where available, the temporal baseline, total observed hours, average observation cadence, and method for determining rotation rates. The samples differ in target selection, observation strategy, and rotation measurement methodology. 
We state the differences between each sample throughout this Section as well as Section \ref{catalog_comparison}.

\begin{deluxetable}{lccccccccc}
\tabletypesize{\scriptsize}
\footnotesize{\small}
\tablecaption{Literature Sources. \label{source}}
\tablewidth{0pt}
\tablehead{
\colhead{Cluster}  &\colhead{Objects}\vspace{-0.2cm}&\colhead{Objects After}&\colhead{Telescope} & \colhead{Baseline} & \colhead{Total Observed } & \colhead{Cadence}& \colhead{Rotation Method} & \colhead{Source}\\
 & \colhead{Cross Matched}& \colhead{Quality Cuts}&\colhead{or Observatory}& \colhead{days} &\colhead{hours}& & & &} 
\startdata
             NGC2362 &                       265 &                         44 &       4m Blanco, CTIO &     14  &             32 hr &     6 min\tablenotemark{\footnotesize a} &  LSSC &     \citet{irwin_2008_ngc2362} \\
             NGC2547 &                       172 &                        147 &          2.2m MPG/ESO &    240  &            100 hr &     7 min\tablenotemark{\footnotesize a} &  LSSC &     \citet{irwin_2008_ngc2547} \\
                 M37 &                       355 &                          0 &              6.5m MMT &     31  &                    - &     3 min\tablenotemark{\footnotesize a} &       AoV &    \citet{hartman2009} \\
                 M50 &                       630 &                         46 &       4m Blanco, CTIO &   $\approx$360  &             95 hr &     6 min\tablenotemark{\footnotesize a} &  LSSC &      \citet{irwin_2009_m50} \\
                 M34 &                        48 &                          6 &              2.5m INT &     10  &             45 hr &   3.5 min\tablenotemark{\footnotesize a} &  LSSC & \citet{irwin_2006_m34} \\
                 M34 &                        41 &                         18 &             0.9m WIYN &    143  &                    - &        1 hour\tablenotemark{\footnotesize a} &               L--S &     \citet{meibom_2011_m34} \\
                 M34 &                        15 &                          5 &   42in Hall Telescope &     17  &                    - &             - &            CLEAN &      \citet{james_2010_m34} \\
            Praesepe &                       489 &                        451 &                    K2 &     75  &              75 d &  29.6 min &               L--S &     \citet{douglas2017} \\
            Pleiades &                       605 &                        533 &                    K2 &     72  &              72 d &  29.6 min &               L--S &     \citet{rebull_2016_pleiades} \\
              Hyades &                       156 &                        122 &                    K2 &     80  &              80 d &  29.4 min &               L--S &    \citet{douglas_2019} \\
      Upper Scorpius &                       866 &                        720 &                    K2 &     82  &              82 d &  29.4 min &               L--S &     \citet{rebull2018usco} \\
         $\rho$ Ophiuchus &                       103 &                         84 &                    K2 &     82  &              82 d &  29.4 min &               L--S &     \citet{rebull2018usco} \\
 Field (PAN-STARSS1) &                       263 &                         78 &      1.8m Pan-STARRS1 &  $\approx$1500  &                    - &        $\approx$1 day\tablenotemark{\footnotesize a} &               L--S &  \citet{kado-fong16} \\
   Field (Evryscope) &                        86 &                         72 &       Evryscope array &   $\approx$900  &                    - &     2 min\tablenotemark{\footnotesize a} &               L--S &     \citet{evryscope_2020} \\
      Field (MEarth) &                       357 &                        284 &          MEarth-North &   $\approx$400  &                    - &    20 min\tablenotemark{\footnotesize a} &  LSSC &     \citet{newton2016} \\
      Field (MEarth) &                       263 &                        221 &          MEarth-South &   $\approx$400  &                    - &    30 min\tablenotemark{\footnotesize a} &  LSSC &     \citet{newton_2018} \\
      Field (Kepler) &                      5380 &                       4337 &                Kepler &  $\approx$1000  &           $\approx$1000 d &  29.4 min \tablenotemark{\footnotesize a} &  Auto &  \citet{mcq2014} \\
      Field (CARMENES) & 129 & 113 & - & - & - & - & L--S & \citet{carmenes_rot_19}\\
\enddata
\tablecomments{A list of literature sources of rotation rates for each sample in the catalog. Field(CARMENES) \citep{carmenes_rot_19}, used light curves from a variety of sources, see Section~\ref{source_field}. Baseline represents the time between first and last observation, Total Observed Hours is time spent observing, Cadence is frequency of observation, and Rotation Method is algorithm used to measure the rotation rate from the light curve (\textbf{LSSC} - Least squares sine curves, effectively equivalent to Lomb--Scargle, \textbf{AoV}- Analysis of Variance, \textbf{L--S} - Lomb--Scargle, \textbf{CLEAN} - CLEAN Algorithm \citep{roberts_1987_clean}, \textbf{Auto} - Autocorrelation function.) }
\tablenotetext{a}{Ground based surveys have dirunal and seasonal gaps in observing. This is the minimum cadence achieved by the survey}
\end{deluxetable}

\subsubsection{Clusters}\label{source_clusters}

The goal of gyrochronology is to present a relationship between the rotation rate of a star of a given color and its age. Clusters provide essential snapshots in time of this rotation--age relationship, as each contains a population of stars with a range of colors that are of the same age \citep[e.g., ][]{cameron_2015}. Clusters therefore serve as essential benchmarks for understanding the evolution in rotation rate of M dwarfs.

The intrinsic faintness of M dwarfs makes it challenging for cluster surveys to include them. For example, only a few early M dwarf rotation periods have been measured in NGC 6811 \citep[1 Gyr;][]{curtis2019,meibom_2011_ngc6811_noM}, NGC 752 \citep[1.4 Gyr;][]{Agueros_2018}, and Ruprecht~147 \citep[2.7 Gyr;][]{curtis_2020}, and none in NGC 6819 \citep[2.5 Gyr;][]{meibom_2015_ngc6819} or M67 \citep[4 Gyr;][]{barnes_2016_m67,esselstein_2018_m67}. The catalog we compiled for this work draws from the cluster surveys that have coverage into the mid M dwarfs and beyond. These include surveys using both ground and space based observatories. We use the results from \citet{rebull2018usco} for Upper Scorpius (USCO) and $\rho$ Ophiuchus, \citet{rebull_2016_pleiades} for Pleiades, \citet{irwin_2009_m50} for M50, \citet{hartman2009} for M37, \citet{irwin_2006_m34}, \citet{james_2010_m34}, \citet{meibom_2011_m34} for M34 (with preference for \citet{meibom_2011_m34} and then \citet{james_2010_m34} in cases where rotation periods differed between sources due to their longer temporal baselines), \citet{irwin_2008_ngc2547} for NGC 2547, \citet{irwin_2008_ngc2362} for NGC2362, \citet{douglas2017} for Praesepe, and \citet{douglas_2019} for the Hyades. 

Objects in associations imaged by K2 (Upper Scorpius, $\rho$ Ophiuchus, Pleiades, and Praesepe, Hyades) had their EPIC designations matched to Gaia DR2 Source IDs in the 1" crossmatch of Gaia DR2 with K2,\footnote{\url{gaia-kepler.fun}, see acknowledgments} with the exception of  Hyades objects, as \citet{douglas_2019} already provided Gaia DR2 Source IDs for them. For the other associations (M50, M37, M34, NGC 2547, NGC 2362) we performed a 2" cross match with Gaia DR2. We visually expected the cross-matched objects on a $V-I$ vs $G$-$G_{\rm RP}$ plot for outliers. Once cross-matched with Gaia DR2, every object in each association was subjected to the astrometric and photometric quality cuts of \citet{kiman_2019}. No objects in M37 passed the quality cuts, and so it was not included in our analysis. The total number of cross matched sources are listed in Table \ref{source} as well as final number after quality cuts. 

\subsubsection{Field Star Samples}\label{source_field}

We also included literature samples in our analysis that do not focus on a single cluster of stars, which we refer to as field star samples. Individual objects are considered to be of unknown ages, but together they provide insight into the nature of M dwarfs across their entire age range of the solar neighborhood.

The largest of these samples is from Kepler \citep{mcq2014}. Aside from the four clusters targeted, the Kepler sample is assumed to be composed of unassociated field stars of unknown ages. We used an existing  cross match\footnote{\url{gaia-kepler.fun}, see acknowledgments} to associate Kepler IDs from the list of periodic rotation rates from \citet{mcq2014} to Gaia Source ID's and measurements. The full Kepler sample extends to higher masses above the M dwarf regime, so following the recommendation of \citet{kiman_2019} we make a cut at Gaia ($G - G_{\rm RP}$) = 0.8 to limit the objects to M0 or later in the sample.

Other field rotation rate samples include MEarth North \citep{newton2016}, MEarth South \citep{newton_2018}, Evryscope \citep{evryscope_2020}, Pan-STARRS1 Medium-Deep Survey (PS1-MDS) \citep{kado-fong16}. These are all ground-based surveys with long ($>1yr$) temporal baselines. The MEarth project targets nearby M dwarfs across the whole sky intending to be sensitive to transits of earth-sized planets around M type hosts. Evryscope observes bright nearby stars in the southern sky, and the rotation sample is based on objects with previously known flaring events \citep{howard_2019_flare}. \citet{evryscope_2020} used Gaia distances and a magnitude cutoff to select late K and M dwarfs. \citet{kado-fong16} used the PS1-MDS which targeted five fields spread out in right ascension. They used various photometry cuts from PS1-MDS to select for M dwarfs. Their diffuse sky positions presumes that they are unassociated field stars of unknown ages. 

Additionally we included rotation rates for objects from the CARMENES project\footnote{https://carmenes.caha.es/}. \citet{carmenes_rot_19} measured rotation rates for from publically available photometric lightcurves (including MEarth, ASAS, SuperWASP, NSVS, Catalina, ASAS-SN, K2, and HATNet) where available as well as producing light curves for a few objects using 0.2-0.8 meter telescopes. For full details see Section 2 in \citet{carmenes_rot_19}. The CARMENES rotation rates are gathered from a range of different observation methods. As our combined field sample is also drawn from a variety surveys, we find it appropriate to include the rotation rates to add valuable additional objects.

For all the samples we performed a 2" cross match with Gaia DR2 to obtain their Gaia photometry and astrometry, and then applied the \citet{kiman_2019} quality cuts. We list the number of objects from each sample that are cross matched to a Gaia DR2 source, and then how many pass the photometry cuts in Table \ref{source}.

\subsubsection{Comparing samples}\label{catalog_comparison}

We show the combined rotation rates and relevant Gaia DR2 data in Table \ref{litrot}. Through the remainder of this section we comment on factors relevant to comparing the rotation rates across the catalogs.

The light curves and photometry from which the rotation rates are derived come from a variety of optical bands. Certain forms of stellar variability are wavelength dependent (e.g., the corona is primarily studied in X-rays), however all the $P_{\rm rot}$ discussed in this work are from optical wavelengths. The amplitudes and time scales of the variability in each sample are considered by their authors to be due to starspot movement and modulation in the photosphere and are therefore related to the rotation rate of the star.

The temporal coverage of the observations define the longest and shortest rotation rates that surveys are effectively sensitive to. The fastest rotation rate for an M dwarf is of a few hours, on the order of the breakup velocity for their size and mass \citep{herbst_07}. With peak cadences between $3-30$ minutes, all of the samples are sensitive to the minimum possible periods for the objects of interest in this work. The longest rotation period measured for an M dwarf comes from MEarth and is over 100 days long. To accurately measure a rotation period, a survey needs to have a temporal baseline on the order of a given period, and preferably greater to confirm it by observing multiple revolutions. Ground-based surveys are limited by observing time constraints from the Sun, Moon, and sky conditions, and space-based missions are limited by instrument systematics or mission parameters (e.g., K2 campaigns only last $\approx$80 days, whereas most Kepler targets were observed for $\approx$4 years), and so most of these surveys are not sensitive to the longest M dwarf rotation rates. Each has their own upper sensitivity in period space based on their baseline. However, there are physical reasons for the maximum period of a population of M dwarfs to differ with their age (see Sections \ref{6rotation_age} and \ref{7halphakin}), and therefore the cluster samples included may well have the requisite maximum temporal sensitivity for their targets even when their baselines are under 100 days. Long baselines are most important for surveys looking at field M dwarf stars, where there are known examples of $\geq$ 100 day rotation periods for mid- to late-Ms.

The method by which the rotation rates are measured differs from survey to survey. The most common method is the Lomb--Scargle algorithm \citep{lomb_1976,scargle_1982,vandJ2015_ls,vander_2018} also referred to as least squares fitting of sine curves \citep[e.g.,][]{irwin_2006_m34}. This method is effective for well sampled time series, but also for those with gaps in observing coverage. At a minimum, ground-based surveys have a diurnal observation schedule, and often have far greater gaps between observations. In the range of Kepler and K2 there were no significant gaps in observation, allowing for other techniques. For example, \citet{mcq2014} used the auto-correlation function (ACF) for the Kepler data set, which requires evenly-spaced data. Lomb--Scargle periodograms are quite effective for active star and \citet{rebull_2016_pleiades}, \citet{rebull2018usco}, \citet{douglas2017} and \citet{douglas_2019} all use this method for their respective K2 target clusters. However, even though they use the same method, there is a difference in their implementation, as \citeauthor{rebull_2016_pleiades} searched for periods in the range of $0.05-35$ days, as compared to $0.1-70$ by \citet{douglas2017}. In \citet{gillen_2019} three different methods (ACF, Lomb--Scargle, and Gaussian Processes) were compared to measure light curves of Blanco 1 members ($\sim$100 Myr), and while there were some objects that had periods disagreeing up to 15\% between methods, most varied by $<$ 2\%, with better agreement for objects $(G-K_{\rm s}) > 2.5$, or spectral type K5 and later. Therefore, while the method for measuring the rotation period varies, we find it appropriate to compare them across our sample, despite their heterogeneous methodologies.

\begin{deluxetable}{lccchchcccchcc}
\tabletypesize{\scriptsize}

\tablecaption{Literature Rotation Catalog. \label{litrot}}
\tablewidth{0pt}
\tablehead{
\colhead{Gaia DR2 Source ID} & \colhead{R.A.} & \colhead{Decl.}& \colhead{pmRA} & \nocolhead{pmRA Error} &  \colhead{pmDec} & \nocolhead{pmDEC Error} &  \colhead{$G$} & \colhead{$G_{\rm BP}$} & \colhead{$G_{\rm RP}$} & \colhead{Parallax} & \nocolhead{Parallax error} & \colhead{Source} & \colhead{$P_{\rm rot}$}\\
 & (deg) & (deg) & (mas yr$^{-1}$) & (mas yr$^{-1}$) & (mas yr$^{-1}$)& (mas yr$^{-1}$) & (mag) & (mag) & (mag) & (arcsec) & (arcsec) & & (day)} 
\startdata
 522864272037653504 &  15.839426 &  62.365881 & 730.740 &     0.163 &  86.352 &      0.225 &      11.924 &       13.907 &       10.603 & 101.637 &    0.081 & carmenes &   1.0200 \\
4761270593953489664 &  77.444871 & $-$60.001702 &  69.101 &     0.042 &  $-$1.247 &      0.051 &      12.553 &       13.371 &       11.685 &   9.274 &    0.026 &     evry &  45.1000 \\
3313662896312488192 &  66.118535 &  16.886057 & 112.135 &     0.161 & $-$25.507 &      0.093 &       7.656 &        7.980 &        7.200 &  21.666 &    0.068 &   hyades &   7.8400 \\
2051011540009372288 & 290.939595 &  37.006425 &   1.068 &     0.049 &  16.681 &      0.057 &      15.092 &       15.783 &       14.290 &   2.522 &    0.027 &   kepler &  34.7170 \\
 337125726965406592 &  40.883177 &  42.642077 &   1.084 &     0.052 &  $-$5.333 &      0.054 &      13.417 &       13.842 &       12.838 &   1.956 &    0.042 &      m34 &   0.8100 \\
3451192967515845120 &  87.918167 &  32.588942 &   1.704 &     0.114 &  $-$5.850 &      0.092 &      16.133 &       16.631 &       15.478 &   0.730 &    0.060 &      m37 &   5.6558 \\
3051571789907383168 & 105.730501 &  $-$8.262112 &  $-$7.687 &     0.163 &   5.950 &      0.149 &      17.337 &       18.028 &       16.485 &   1.063 &    0.098 &      m50 &   7.3580 \\
2779735819320035840 &   9.892268 &  14.909837 & 322.341 &     0.249 &  24.722 &      0.161 &      12.984 &       14.778 &       11.699 &  35.609 &    0.158 &   mearth &   1.5920 \\
5616921778334790144 & 109.908838 & $-$24.713436 &  $-$2.629 &     0.062 &   2.681 &      0.084 &      16.308 &       17.101 &       15.429 &   0.741 &    0.048 &  ngc2362 &   5.9990 \\
5514364861310955008 & 122.347993 & $-$49.274202 &  $-$9.036 &     0.379 &   3.619 &      0.339 &      18.885 &       20.294 &       17.515 &   2.451 &    0.196 &  ngc2547 &   0.9430 \\
  50503115981677056 &  59.819400 &  19.740776 &  13.396 &     0.222 & $-$40.106 &      0.120 &      16.701 &       18.394 &       15.425 &   6.233 &    0.113 &     plei &   0.3727 \\
 658156527832150400 & 131.048690 &  16.542345 & $-$36.876 &     0.232 & $-$10.673 &      0.133 &      17.604 &       19.334 &       16.303 &   5.010 &    0.141 &     prae &   1.0400 \\
2489945648485976320 &  36.377600 &  $-$4.584857 &   9.132 &     0.320 & $-$18.904 &      0.280 &      18.094 &       19.434 &       16.881 &   2.932 &    0.166 &  ps1-mds & 120.9684 \\
6049049811047121536 & 247.080079 & $-$24.959581 &  $-$5.807 &     0.202 & $-$22.795 &      0.143 &      14.946 &       16.965 &       13.591 &   6.926 &    0.100 &     roph &  11.9840 \\
6038892694440227712 & 244.483638 & $-$28.944536 & $-$10.657 &     0.356 & $-$21.362 &      0.244 &      16.393 &       18.546 &       15.007 &   6.589 &    0.116 &     usco &   1.3723 \\
\enddata
\tablecomments{Gaia DR2 data for 1 object from each group, a sample from the 8296 row table in the full online version. (Prot) values come from their respective sources. (Source) column lists membership as so \textbf{NGC2362} - NGC2362, \textbf{NGC2547} - NGC2547, \textbf{m34} - M34, \textbf{m37} - M37,\textbf{m50} - M50,\textbf{hyades} - Hyades,\textbf{plei} - Pleiades,\textbf{prae} - Praesepe,\textbf{roph} - $\rho$ Ophiuchus,\textbf{usco} - Upper Scorpius}
\end{deluxetable}

\subsection{Correcting Extinction in Gaia DR2 Photometry}

While Gaia DR2 provides a uniform source for the photometry, the galactic environment and amount of reddening due to interstellar dust varies from object to object depending on their 3D position in the solar neighborhood. We follow the methodology from \citet{gagne_mutau_2021} to calculate extinction corrected ($G$-$G_{\rm RP}$) color for every object in K2SDSS and our literature cluster and field star samples. The method is based on STILISM \citep{lallement_2014,capitanio_2017,lallement_2018}\footnote{Available at https://stilism.obspm.fr}, and determines extinction values for individual objects based on their sky position, Gaia DR2 distance and a photometric spectral type estimate. The extinction corrected ($G$-$G_{\rm RP}$) color is then used to obtain a better photometric spectral type, and the process is repeated until there is no change in spectral type. Calculated extinction coefficients and corrected Gaia DR2 photometry for every object are presented in Table~\ref{table:corrected_phot}. We use the corrected photometry throughout the rest of the paper.

\begin{deluxetable}{lccccccccc}
\tabletypesize{\scriptsize}

\tablecaption{Corrected Gaia DR2 Photometry. \label{table:corrected_phot}}
\tablewidth{0pt}
\tablehead{
\colhead{Gaia DR2 Source ID} & \colhead{$E(B-V)$} & \colhead{Error $E(B-V)$}& \colhead{$A{\rm G}$} & \colhead{$A{\rm BP}$} &  \colhead{$A{\rm RP}$} & \colhead{$G$ Corrected} &  \colhead{$G_{\rm BP}$ Corrected} & \colhead{$G_{\rm RP}$ Corrected}\\
 & (mag) & (mag) & (mag) & (mag) & (mag)& (mag) & (mag) & (mag)} 
\startdata
64917094946846848 &   0.048 &       0.017 & 0.090 & 0.140 & 0.080 &  16.011 &   17.586 &   14.781 \\
64921458633614976 &   0.052 &       0.017 & 0.130 & 0.166 & 0.096 &  11.192 &   11.645 &   10.619 \\
64922283267331968 &   0.047 &       0.017 & 0.105 & 0.141 & 0.085 &  13.551 &   14.411 &   12.643 \\
64923279699744256 &   0.050 &       0.017 & 0.125 & 0.160 & 0.093 &  11.345 &   11.781 &   10.785 \\
64924413571101952 &   0.056 &       0.018 & 0.146 & 0.184 & 0.104 &   9.989 &   10.277 &    9.571 \\
64925444363253632 &   0.040 &       0.017 & 0.068 & 0.115 & 0.063 &  17.570 &   19.555 &   16.107 \\
64927368508592512 &   0.046 &       0.016 & 0.084 & 0.135 & 0.076 &  17.478 &   19.272 &   16.157 \\
64928605459180416 &   0.050 &       0.017 & 0.128 & 0.163 & 0.093 &  10.801 &   11.186 &   10.299 \\
64928674178659968 &   0.050 &       0.017 & 0.092 & 0.147 & 0.083 &  17.458 &   19.205 &   16.147 \\
64930495244783616 &   0.049 &       0.017 & 0.134 & 0.166 & 0.093 &   8.893 &    9.101 &    8.578 \\
64933037865353600 &   0.007 &       0.015 & 0.012 & 0.020 & 0.011 &  15.638 &   17.702 &   14.280 \\
64933621980897536 &   0.035 &       0.015 & 0.066 & 0.102 & 0.058 &  16.036 &   17.813 &   14.740 \\
64933759417769984 &   0.049 &       0.017 & 0.138 & 0.167 & 0.093 &   8.186 &    8.306 &    7.989 \\
64934893291213952 &   0.046 &       0.016 & 0.107 & 0.141 & 0.084 &  12.140 &   12.711 &   11.435 \\
64936508198907904 &   0.046 &       0.017 & 0.106 & 0.140 & 0.084 &  12.789 &   13.469 &   12.006 \\
64938569783224960 &   0.060 &       0.019 & 0.119 & 0.177 & 0.102 &  15.750 &   17.193 &   14.567 \\
64941211186036352 &   0.047 &       0.017 & 0.122 & 0.154 & 0.088 &  10.089 &   10.388 &    9.658 \\
64942001460286080 &   0.046 &       0.017 & 0.084 & 0.134 & 0.075 &  17.629 &   19.591 &   16.263 \\
64943758103791104 &   0.045 &       0.016 & 0.086 & 0.133 & 0.076 &  16.326 &   18.096 &   15.032 \\
64944170420647296 &   0.057 &       0.019 & 0.125 & 0.172 & 0.102 &  13.457 &   14.333 &   12.505 \\
\enddata
\tablecomments{Gaia DR2 photometry corrected for extinction for each object in our total sample. This an example of the 8948 row table in the full online version.}
\end{deluxetable}

%%%%%%%%%%%%%%%%%%%%%%%%%%%%%%%%%%%%%%%%%%%%%%%%%%%%%%%%%%%%%%%%%%%%%%%%%%%%%%%%%%%%%%%%%%%%%%%%%%%%%%%%%%%%%%%%%%%%%%%%
\section{Measuring new Rotation periods for the K2SDSS sample}\label{sec:rotation_k2sdss}

The K2SDSS sample introduced in section \ref{k2sdssintro} consists of \ksdssobjs objects which were targeted by at least one campaign during the K2 mission. We detail our methods for extracting rotation periods from the light curves of targets from those campaigns in this section. Additionally, we comment on our recovery rate in Section \ref{rotation_recovery}, as well as objects with dynamic light curves in Section \ref{rotation_interesting}.

We used the EVEREST light curve data products \citep{Luger2018} to measure rotation periods. We considered other detrending packages such as K2SFF \citep{VandJ2014} and K2SC \citep{Aigrain2016}, but chose EVEREST as we had access to products from all campaigns and found that it still maintained the astrophysical signals after detrending.

We used the Lomb--Scargle algorithm as implemented in the \texttt{exoplanet} package \citep{exoplanet} to identify likely rotation periods, and used a bootstrap method to define a significant signal threshold. For each light curve, we randomly shuffled the time series 100 times and returned the maximum power in the Lomb--Scargle periodogram each time. We required a period from an observed light curve to have a power greater than 99\% of the bootstrap periods to consider it significant. We searched for periods of 0.05 to 45 days, and applied a high pass filter at periods beyond half the observing window for each campaign. This suppressed erroneous power attributed to long term trends left over from detrending.

\begin{figure}
\centering
\begin{minipage}{.45\textwidth}
  \includegraphics[width=\linewidth]{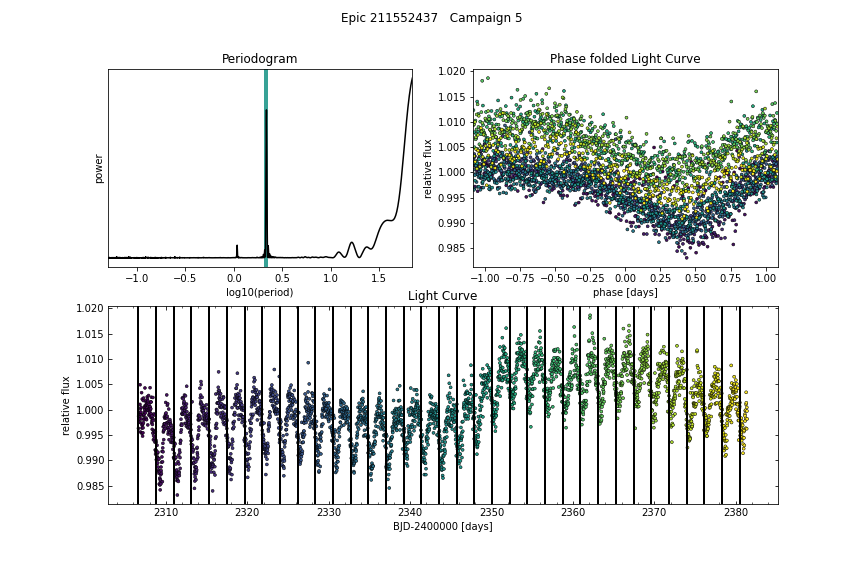}
    \caption{This light curve received a score of 0. It shows a clear and repeated amplitude variation.}
    \label{fig:0score}
\end{minipage}
\quad
\begin{minipage}{.45\textwidth}
  \includegraphics[width=\linewidth]{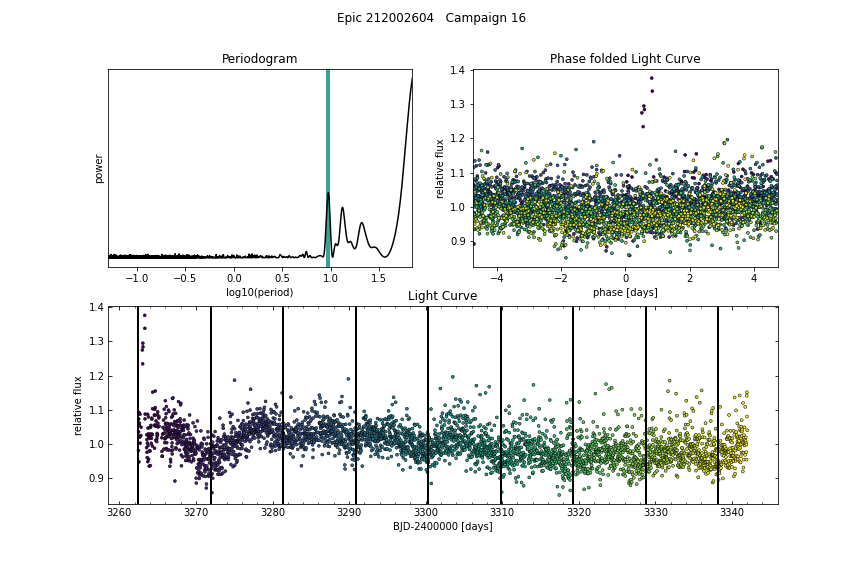}
    \caption{This light curve received a score of 1. While there is amplitude variation, it is not as clear.}
    \label{fig:1score}
\end{minipage}
\end{figure}

\begin{figure}
\centering
\begin{minipage}{.45\textwidth}
  \includegraphics[width=\linewidth]{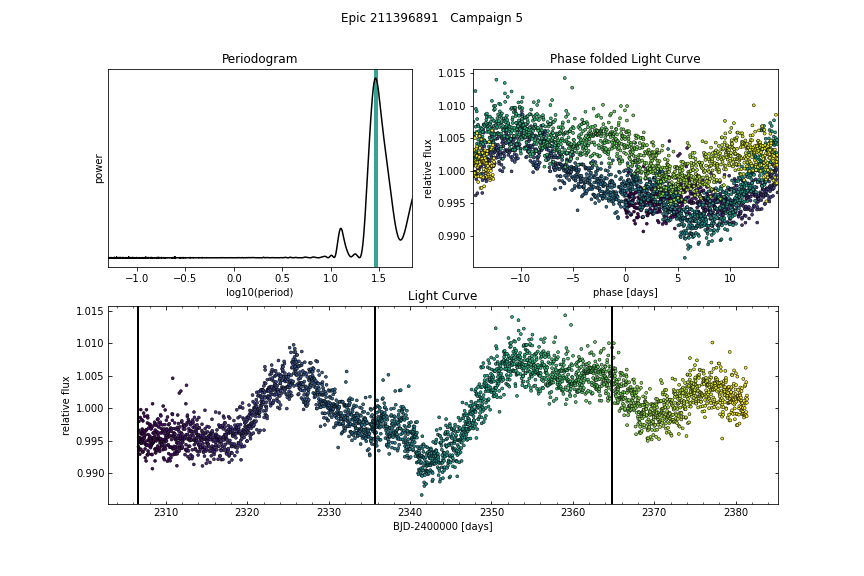}
    \caption{This light curve received a score of 2. While there appears to be variation due to rotation, it is not sure enough to be included in the final rotation period list.}
    \label{fig:2score}
\end{minipage}
\quad
\begin{minipage}{.45\textwidth}
  \includegraphics[width=\linewidth]{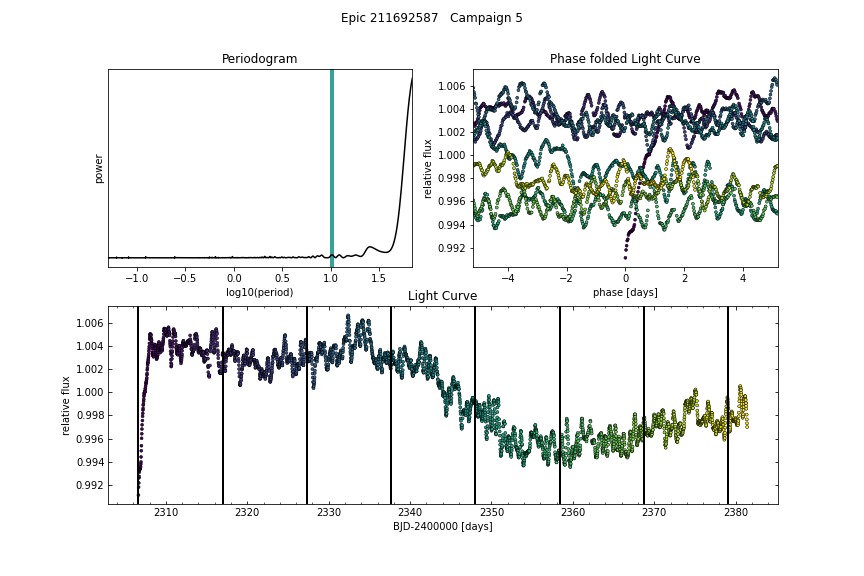}
    \caption{This light curve received a score of 3. While no clear rotation period is seen, the variation is likely astrophysical. See Section \ref{rotation_interesting} for more details.}
    \label{fig:3score}
\end{minipage}
\end{figure}

\begin{figure}
    \centering
    \includegraphics[width=.45\textwidth]{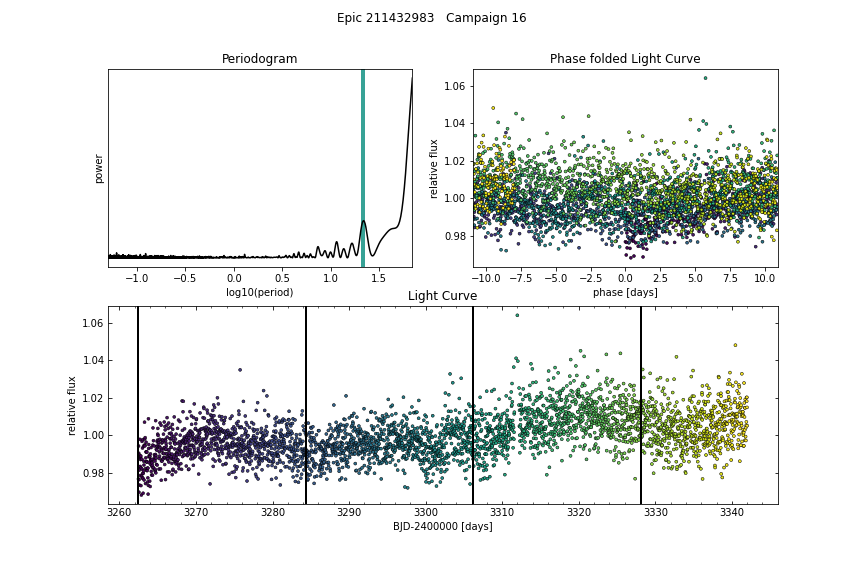}
    \caption{This light curve received a score of 9, and is considered to have no detectable rotation period.}
    \label{fig:9score}
\end{figure}

We visually inspected the full light curve of each target, as well as its periodogram, and a light curve that was phase folded on the period of greatest power. If an object was imaged in multiple K2 campaigns, each light curve was considered separately. Each light curve was then visually scored according to the quality and strength of the rotation period signal. A score of zero was attributed to light curves with clear and obvious periodicities. A score of one was awarded to light curves with strong evidence of periodicity. A score of two was awarded to those with a potential signature of rotation but where the signature was either only present in some of the light curves, seemed to have a cycle longer than the campaign, or was below the significance threshold. A score of three was given to objects with peculiar light curves, and a score of nine to those with no evidence of rotational variability. Figures \ref{fig:0score} -- \ref{fig:9score} present examples for each scoring category. We discuss the objects with a score of 3 in Section \ref{rotation_interesting}. We considered light curves with a visual score of 0 or 1 to represent the highest-quality signals in our sample. These have rotation rates listed in the $P_{\rm rot}$ column in Table \ref{table:measuresample}.

\begin{deluxetable}{lcccccc}
\tabletypesize{\scriptsize}

\tablecaption{Measured Rotation Rates for Light Curves of Objects in K2SDSS.\label{table:measuresample}}
\tablewidth{0pt}
\tablehead{
\colhead{EPIC Number} & \colhead{K2 Campaign}&  \colhead{Per 1} & \colhead{Per 2} &\colhead{Quality Flag} & \colhead{Notes}\\
& & (day) & (day) & } 
\startdata
   247991214 &        13 &   3.137 &   1.711 &             0 &              irreg \\
   247968420 &        13 &   2.916 &  15.221 &             0 &              irreg \\
   211920710 &        16 &   1.333 &   0.704 &             0 &             dd \\
   211920710 &         5 &   1.332 &   2.408 &             0 &             dd \\
   211950716 &        16 &   0.245 &   0.124 &             1 &    dd, evo \\
   211744621 &        16 &  16.175 &   6.053 &             2 &       dd \\
   211729756 &        18 &   1.196 &  45.947 &             0 &  dd \\
   211796503 &        18 &   0.267 &   0.136 &             0 &      dd\\
   211729756 &         5 &   2.392 &   0.132 &             0 &       dd \\
   212113915 &        18 &  25.587 &   2.515 &             2 &               \\
   212113915 &         5 &  16.587 &  16.587 &             2 &               \\
   211975927 &         5 &  32.396 &   0.124 &             1 &               \\
   211884659 &         5 &   4.497 &  11.930 &             1 &               \\
   248651399 &        14 &  26.733 &   0.743 &             1 &               \\
   212021699 &         5 &  24.694 &  24.694 &             2 &               \\
   248767646 &        14 &   2.440 &   4.881 &             1 &               \\
   211719815 &        18 &   0.101 &  30.502 &             0 &               \\
   211958609 &        18 &  54.880 &   0.953 &             1 &               \\
   212091105 &        18 &  37.201 &   1.313 &             1 &               \\
   248718280 &        14 &  16.395 &  31.138 &             1 &                \\
\enddata
\tablecomments{Reported periods for objects in K2SDSS. (Per 1) and (Per 2) are the strongest and second strongest periods reported by Lomb--Scargle algorithm. (Quality Flags) and (Notes) are assigned after visual inspection. irreg- irregular variation, dd- double dip light curve, evo- evolution of the lightcurve morphology. See Section~\ref{rotation_interesting}.}
\end{deluxetable}

We check the apertures used in the EVEREST pipeline for contamination from multiple sources for all objects which we report a rotation rate for. We track objects that have another star in the aperture that is unlikely to provide substantial flux, as well as those where there is likely contamination. Of our \ksdssrots objects only 7 had any kind of contamination.

Our final list of rotation rates are provided in Table \ref{table:finalsample}. We relied on visual inspection to choose the final period for objects with light curves from multiple campaigns, but we also list the periods for each campaign in Table \ref{table:measuresample}.

\begin{deluxetable}{lccccc}
\tabletypesize{\scriptsize}

\tablecaption{Final Rotation Rates for Objects in K2SDSS Sample.\label{table:finalsample}}
\tablewidth{0pt}
\tablehead{
\colhead{Epic Number} & \colhead{Period}& \colhead{Quality Flag} &  \colhead{Aperture Flag} & \colhead{Binary Flag}\\
& (day) & & & } 
\startdata
   201384292 &  17.671 &             1 &         0 &       0 \\
   201405570 &   0.555 &             0 &         0 &       0 \\
   201412115 &   2.801 &             0 &         0 &       0 \\
   201412367 &   0.801 &             0 &         0 &       0 \\
   201728540 &   0.948 &             1 &         0 &       0 \\
   201785646 &   4.462 &             0 &         0 &       0 \\
   202065179 &   0.937 &             0 &         0 &       1 \\
   206181579 &   0.437 &             0 &         0 &       0 \\
   206191372 &  17.514 &             0 &         0 &       0 \\
   210674207 &   1.054 &             0 &         0 &       1 \\
   211349910 &   1.247 &             0 &         0 &       0 \\
   211352136 &   0.219 &             0 &         0 &       0 \\
   211355470 &  26.440 &             1 &         0 &       0 \\
   211381169 &   0.103 &             0 &         0 &       0 \\
   211394834 &   1.102 &             0 &         0 &       0 \\
   211422201 &   1.049 &             0 &         0 &       0 \\
   211436693 &   4.306 &             0 &         0 &       0 \\
   211483344 &   1.989 &             0 &         0 &       0 \\
   211490371 &   1.271 &             0 &         0 &       0 \\
   211496911 &  12.542 &             0 &         0 &       0 \\
\enddata
\tablecomments{Final periods for objects in K2SDSS. (Period) is intended to be the reported period, except in cases where (Binary Flag) is not 0, in which case multiple periods were found to be likely stellar signals. See Section~\ref{sec:binaries} and Table~\ref{table:binary}. (Quality Flags) are from visual inspection of the light curve, and are described in Section \ref{sec:rotation_k2sdss}. An (Aperture Flag) value of 1 is possible contamination due to other stars in the light curve, 2 is likely contamination. This an example of the full \ksdssrots row table in the full online version.}
\end{deluxetable}

\subsection{Stars with inconclusive or null period detections}\label{rotation_recovery}

\begin{figure}
    \centering
    \includegraphics[width=5in]{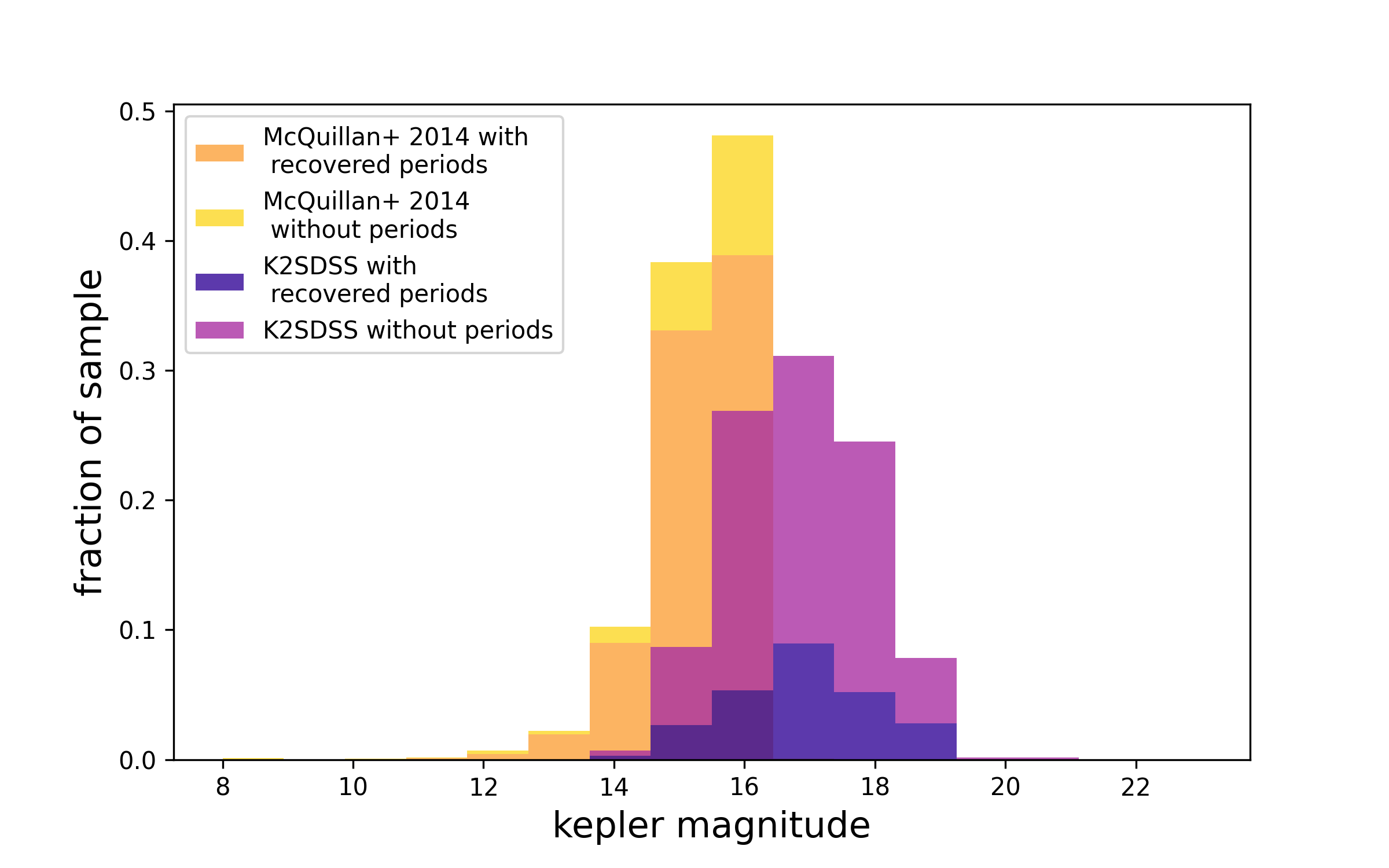}
    \caption{Kepler/K2 magnitudes of the sample fraction with recovered periods and otherwise for \citep{mcq2014} and this work.}
    \label{fig:recovery_hist}
\end{figure}

Table \ref{table:finalsample} presents periods for objects that we are confident we are measuring rotation periods due to stellar variability. The table contains rotation rates for \ksdssrots  of the \ksdssobjs  objects in our original sample.

For comparison, \citet{mcq_mdwarf} successfully measured a rotation rate for 63\% of their 2483 M dwarfs sample (defined by a temperature cut of $<$4000 K in the Kepler Input Catalog) observed in the Kepler field. While temperatures from the KIC can be spurious and the sample may not exclusively consist of M dwarfs, this is still a large discrepancy from our 34 \% recovery rate, which we attribute to several factors. Foremost, many of the objects in K2SDSS are fainter than those attempted in \citet{mcq_mdwarf} whose median Kepler magnitude was 15.44 compared to the median for the K2SDSS sample of 16.92 (see Figure \ref{fig:recovery_hist}). Observing these fainter objects pushes the signal of stellar variability closer to instrumental limitations, which are worse in the K2 mission. The detrending of the increased systematics due to the motion of the spacecraft over each campaign potentially removed or obscured evidence of stellar variability. Additionally, our rotation period measurement method is not sensitive to periods larger than $\sim$40 days, as we required two full, significant rotations to confirm a rotation rate, with the average length of a K2 campaign being around 80 days. However, only 16\% of the rotation rates of objects with $T_{\rm eff} < 4000$ K in \citet{mcq2014} are longer than 40 days, so this does not account for the whole difference if we assume that the rotation period distributions are similar. With these caveats in mind, we do not find a reason for the discrepancy in recovery rate between the K2SDSS sample and the M dwarfs in \citet{mcq2014}.

\subsection{Peculiar objects}\label{rotation_interesting}
The process of this work included visually inspecting 1895 light curves. In this section, we present the peculiar objects that require further notes (including those with a score of 3 from Section \ref{sec:rotation_k2sdss}).

\subsubsection{Double dip, irregular, or rapidly evolving}

The distribution of starspot positions on the surface of a star determines the pattern of stellar variability captured in the light curve of an object \citep{berdy_review}. We follow the example of works such as \citet{rebull_2016_pleiades} and \citet{douglas2017} and identify light curves with distinctive features. For example, multiple starspots separated across opposing hemispheres can create two local minima in brightness within a single rotation cycle, or a double dip. This can often return two peaks in the Lomb--Scargle periodogram, one being a harmonic or half harmonic of a period along with the true period. We identified these light curves with a 'dd' in the notes column of Table \ref{table:finalsample}.

We also flag light curves with changes in their light morphologies over the cycles in a light curve. While the period is stable, the shape of the light curve changes between cycles in some combination of magnitude or position of maximum or minimum in phase. As this is potentially due to starspot evolution, these objects have an 'evo' in the note section for the light curve.

Additionally a few objects have irregular cycles of variability or semi-periodic cycles. The variation of the light curve may still be caused by stellar variability, but it is more likely explained by an occulting circumstellar disk (eg. the dippers in \citet{rebull2018usco}). These have an 'irreg' in the notes column of Table \ref{table:finalsample}.

\subsubsection{Potential Binaries}\label{sec:binaries}
In several of the light curves in our sample there are two peaks in the periodograms that are not integer multiples of each other. Unlike periodograms of double dipping stars, these light curves cannot be described with harmonics. \citep{stauffer_bin_18} identified similar multi-period K2 lightcurves of M dwarfs as binary systems, and so we list the lightcurves in our sample as potential binaries. We verified that each period was a legitimate signal of stellar rotation by visually inspecting and phase folding the light curves on both signals. We present the EPIC numbers and periods for these objects in Table~\ref{table:binary} as potential binaries.

\begin{deluxetable}{lccccc}
\tabletypesize{\scriptsize}

\tablecaption{Potential Binaries in K2SDSS.\label{table:binary}}
\tablewidth{0pt}
\tablehead{
\colhead{EPIC number} &  \colhead{Period 1} & \colhead{Period 2} &\colhead{Beat Flag} & \colhead{Gaia DR2 $RUWE$}\\
& (day) & (day) & & \\}
\startdata
   202065179 &     0.937 &     0.642 &          0 & 1.030\\
   210674207 &     1.054 &     0.987 &          1 & 1.15\\
   211616100 &     0.307 &     0.251 &          0 & 0.96\\
   211707676 &     0.309 &     0.399 &          0 & 0.995\\
   211942008 &    22.625 &     2.750 &          0 & 1.37\\
   211980019 &    25.713 &     1.050 &          0 & 1.587\\
   211981954 &     0.280 &     0.411 &          0 & 0.977\\
   212019609 &     0.779 &     3.366 &          0 & 1.2199\\
   212148445 &     4.872 &     5.400 &          1 & 0.981\\
   220191843 &     4.443 &     3.873 &          1 & 1.023\\
\enddata
\tablecomments{Potential binaries in K2SDSS. (Per 1) and (Per 2) are both considered potential stellar rotation signatures. (Beat Flag) tracks light curves in which a beat pattern is observed in the light curve. (Gaia DR2 RUWE) is an indication of excess astrometric noise in a single star solution for Gaia DR2, with larger values indicating the presence of a companion.}
\end{deluxetable}

Of our potential binaries only EPIC 212148445 has potential contamination from another source in its EVEREST aperture (See Table~\ref{table:finalsample}). The other source in its aperture has no Gaia DR2 Source ID, therefore is likely close to the Gaia limiting magnitude of $G\approx$21 and its contribution to the flux measured in the light curve is probably negligible.  

We check the $RUWE$ from Gaia DR 2 for our potential binaries. A $RUWE$ $\geq$ 1.4 is one line of evidence that a binary companion might be effecting the astrometric solution for an object. Of our potential binaries, EPIC 211980019 has a $RUWE$ = 1.587 while EPIC 211942008 has a $RUWE$ = 1.37, as well as another Gaia DR2 Source within 0.762". We interpret these as likely being wide binaries. The rest of the potential binaries do not have significant $RUWE$ values, however more recent work \citep{ruwe_2021} has suggested that even a $RUWE$ $>$ 1.0 could be indicative of a companion. Only 4 of our binary candidates are below this threshold, with EPIC 211616100 having the smallest value of $RUWE$ = 0.96.

\subsubsection{Beat Period Binaries}\label{sec:beat_binaries}

A subset of the light curves with two strong peaks in the periodogram have two peaks closely spaced together in frequency space. The difference in the frequencies creates a beat pattern on the total light curve. They are flagged with a value of 1 in the 'beat flag' column of Table~\ref{table:binary}. \citet{rebull2018usco} keeps tracks of these beat signals as well, finding 107 in Upper Scorpius, and 10 in $\rho$ Ophiuchus. Furthermore, \citet{paudel2019} present evidence that the light curve for EPIC 248624299 which has a beat pattern is a binary system of M dwarfs with very similar but distinct rotation rates which when combined created an additional beat frequency in the light curve of the object. We treat this as an example of a beat binary system, of which we flagged 3 additional examples in our sample. While we are labeling them as candidate binary systems, it is important to note that this type of light curve could also potentially be due to differential rotation of sunspots at distinct latitudes on the surface of the star.

\begin{figure}
    \centering
    \includegraphics[width=5in]{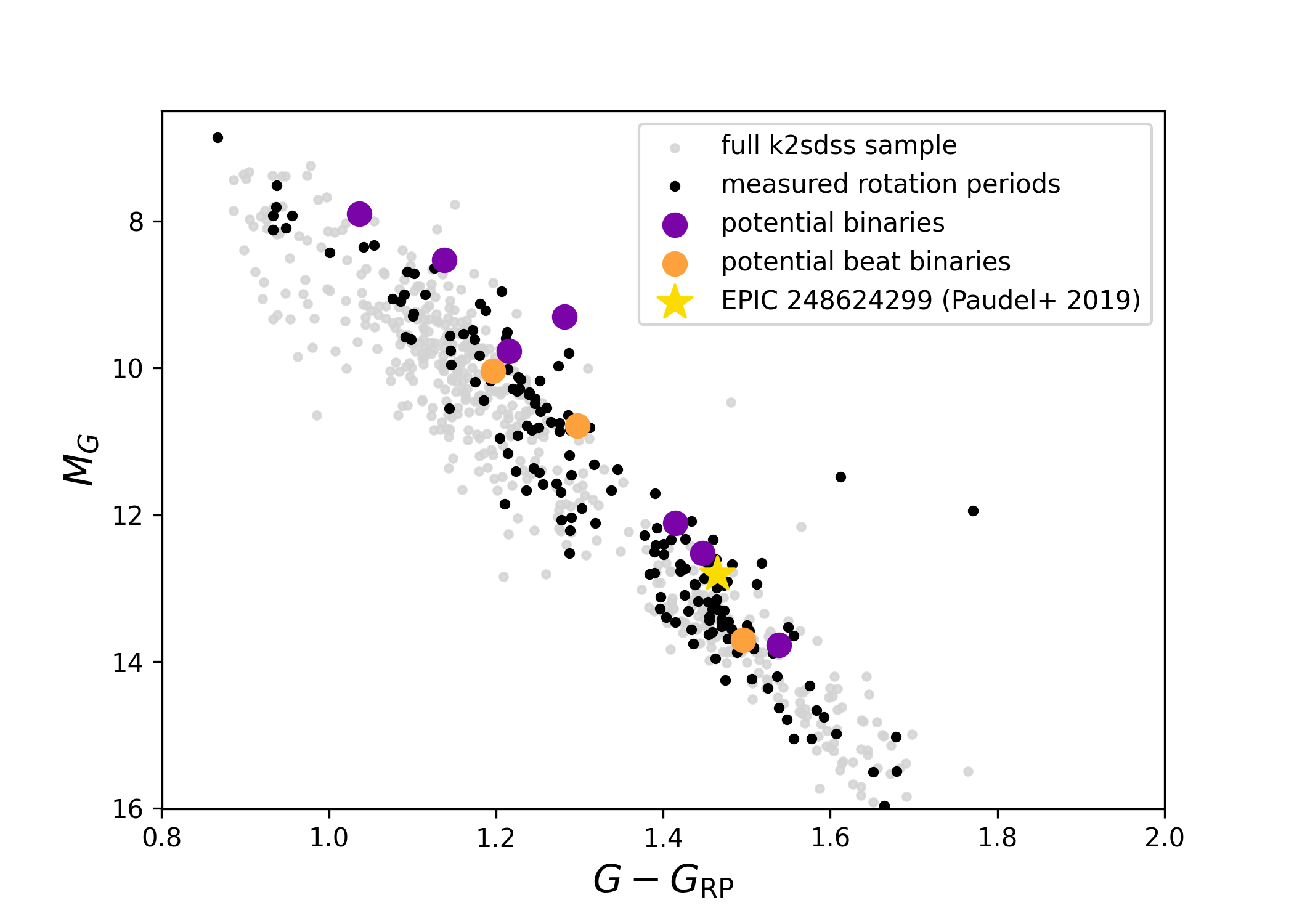}
    \caption{A color--magnitude diagram of the K2SDSS sample, distinguishing those with recovered periods and highlighting candidate binaries based on detection of multiple periods. EPIC 248624299, shown as a star, is an example of a binary discovered from its multiple periods \citep{paudel2019}. Candidate binaries tend to fall above the rest of the sample as expected. The gap at $G - G_{\rm RP}$ = 1.35 is a selection effect leftover from SDSS.}
    \label{fig:binaries}
\end{figure}

Furthermore, we compare the photometric distances from \citet{west2011} (derived using the $M_{\rm r}$, $r$--$z$ parallax relation from \citet{bochanski_2010}) for the objects in Table \ref{table:binary} to the distances derived from inversion of the Gaia DR2 parallax measurements. In all cases, the photometric distances are underestimated and off by over 5$\sigma$ compared to their trigonometric distance. This disparity implies binarity, with the companion's additional flux causing the photometric distance to be underestimated. We place all the binary candidates on a color--magnitude diagram along with the rest of the K2SDSS sample in Figure \ref{fig:binaries}. This overluminousity of the objects places them on the upper edge of the stellar sequence compared to the rest of the K2SDSS field stars, which is more evidence that they are likely binary systems. However, without multi-epoch radial velocity measurements, we cannot know their periods or separations.

%%%%%%%%%%%%%%%%%%%%%%%%%%%%%%%%%%%%%%%%%%%%%%%%%%%%%%%%%%%%%%%%%%%%%%%%%%%%%%%%%%%%%%%%%%%%%%%%%%%%%%%%%%%%%%%%%%%%%%%%
\section{The M dwarf Rotation Sample on A Gaia DR2 Color Magnitude Diagram}\label{5rotation_cmd}

This section presents the Gaia DR2 CMD for M dwarfs with measured rotation rates, which places these objects from various sources in a larger uniform context, one that informs our understanding of the rotation--age relation. The literature sample spans decades of observations at various facilities including space-based missions and ground-based telescopes of varying size. Gaia DR2 provides us with data on a common astrometric and photometric system with which all of the stars in our sample have been observed, and removes the need to convert photometry between the differing filters. The Gaia DR2 catalog covers the entire sky and yields high-precision astrometric measurements with uniform photometry. Therefore the Gaia DR2 catalog provides a powerful analysis tool for a disparately collected rotation sample.

The CMD is plotted on axes of the absolute magnitude in $G$ band ($M_{\rm G}$) which was calculated from the mean $G$ band photometry and the inverse parallax measurement from Gaia DR2, as well as the mean $G$ magnitude minus the mean $G_{\rm RP}$ magnitude ($G - G_{\rm RP}$). The ($G - G_{\rm RP}$) color has been shown to have the tightest relation with spectral type for M dwarfs when using Gaia photometry \citep{kiman_2019}. 

Figure~\ref{fig:cmd_age_wk2sdss} shows our literature sample in a Gaia CMD, with points colored by the log of their reported age, youngest as yellow and oldest as dark purple. Field objects with rotation rates but unknown ages are shown in grey. Additionally, 23,842 M dwarfs without rotations from the MLSDSS--GaiaDR2 sample are presented as a gray scale density map and contours (see \citep{kiman_2019}). While these individual objects do not have rotation rates, we show their positions to indicate the spread of field M dwarfs. 

Members of different clusters fall along distinct sequences in Figure~\ref{fig:cmd_age_wk2sdss}, where young clusters with ages below 100 Myr are shifted to brighter $M_G$ magnitudes. This is interpreted as these stars still being in the process of contracting in radius and fully settling on to the main sequence \citep{mann_2015}. Younger clusters also tend to have more scatter in their main sequences than older ones. For example, objects at ($G - G_{\rm RP}$) = 1.3 in the Upper Scorpius association (orange points) have a $M_G$ range of $\sim$1 mag compared to Praesepe members which only have a 0.5 mag scatter. The individual older clusters (ages $>$100 Myr, light and dark purple points) are difficult to distinguish as they overlap in position on the CMD. Some stars from the presumed field sample overlap with the oldest clusters, however many field stars are fainter or bluer than most on a cluster stellar sequence, consistent with those objects in the field sample being in a range of ages older than 1 Gyr. The majority of field objects fall within the contours of the MLSDSS--GaiaDR2 sample. Young associations are located on the upper edge of the filled sequence, and the youngest associations lie above it.

While the ($G - G_{\rm RP}$) color is used as a common property to compare these objects, it is important to note its limitations. Color is a proxy of temperature, which for main-sequence stars is associated with their mass. However, this is not a one to one correlation for the coolest stars. Evolutionary models of low-mass stars \citep[e.g.,][]{baraffe} reveal that there is a significant change in the temperature and brightness of late M dwarfs over the first several Myr at a fixed mass. Naively, taking a slice in the color axis where objects of different ages are plotted would often result in comparison of objects of different masses that happen to have the same color or temperature during their evolution, especially for late-type M dwarfs. It is therefore important to be careful when directly comparing color bins between objects that are known to be younger than $10-40$\,Myr and those of older objects.

Furthermore, the area ($G- G_{\rm RP}) \geq $1.4 in Figure~\ref{fig:cmd_age_wk2sdss} could suffer contamination due to brown dwarfs. The hydrogen burning limit delineates a star from a brown dwarf, roughly at 75 $M_{\rm Jup}$ \citep{kumar}. Despite being less massive, brown dwarfs begin with temperatures as warm as those of late M dwarfs before cooling over their lifetimes. The $G - G_{\rm RP}$ color is a good proxy for temperature, and therefore mass and spectral type, for M dwarfs, but cannot distinguish true M dwarfs from young and warm brown dwarfs. Since brown dwarfs cool over time, the color at which this contamination needs to be considered depends on age. It is therefore important to consider contamination by brown dwarfs especially in the young samples. The substellar boundary on the ($G - G_{\rm RP}$) axis moves redward in time across the cluster samples. Temperature versus spectral types from \citet{filippazzo_2014}, updated in \citet{faherty_2016}, and evolutionary models from \citet{saumon_marley_2008} estimate that for the 10 Myr old Upper Scorpius association, the substellar boundary corresponds to spectral type $\sim$M6, whereas for the Pleiades at 120 Myr corresponds to $\sim$M8. For the purposes of this work, brown dwarf contamination on the CMD of Figure~\ref{fig:cmd_age_wk2sdss} is negligible for ages older than that of the Pleiades. 

\subsection{The K2SDSS sample on the Gaia CMD}

\begin{figure}
    \centering
    \includegraphics[width=\textwidth]{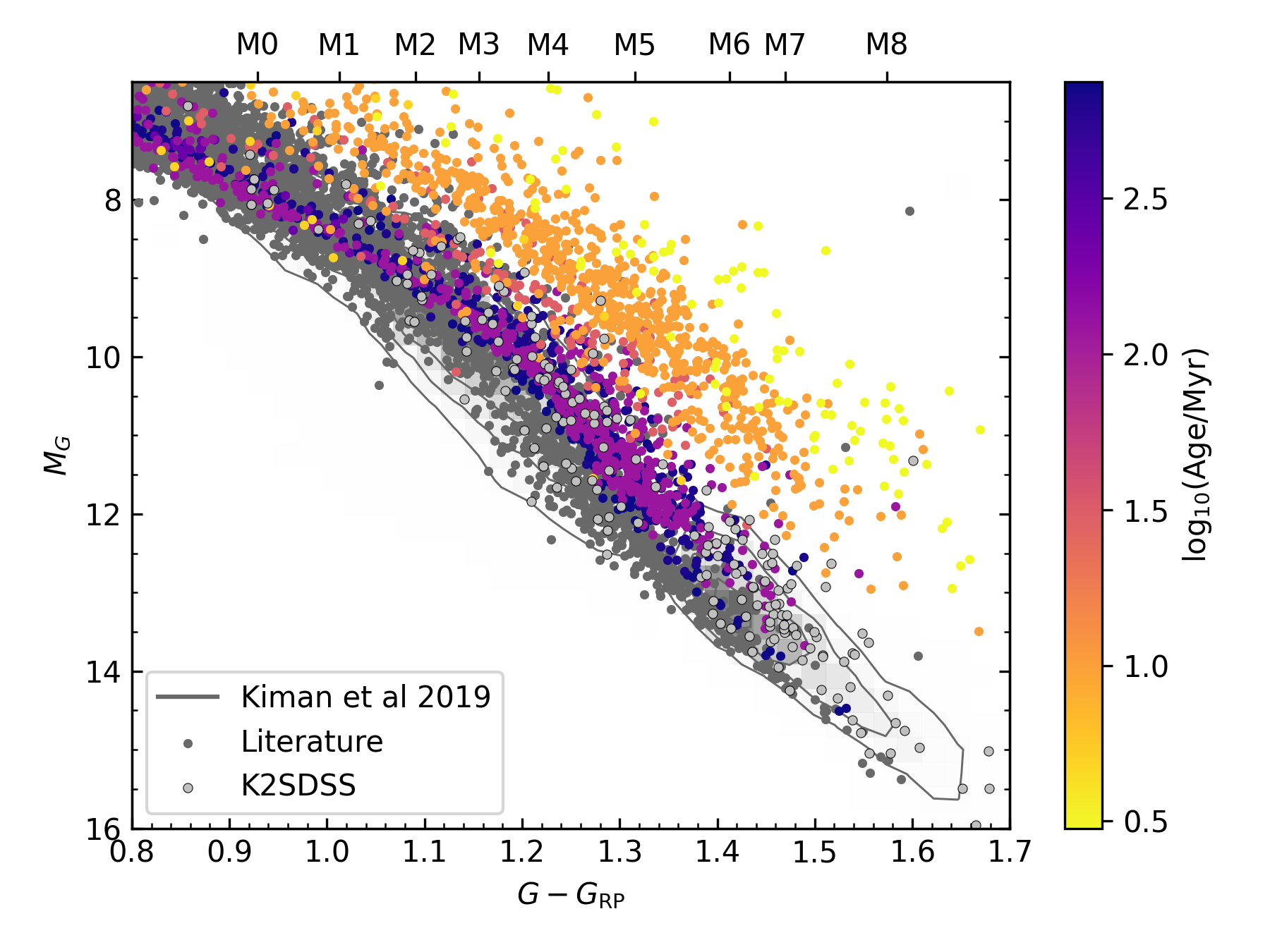}
    \caption{A CMD color coded by age with clusters and field samples with K2SDSS over-plotted. Clusters are color coded by log(age), while field objects are shown in gray. The K2SDSS sample adds significantly more old objects beyond $G - G_{\rm RP}$ = 1.4.}
    \label{fig:cmd_age_wk2sdss}
\end{figure}

The \ksdssrots objects in the K2SDSS sample with rotation rates measured for the first time in this work are light grey filled circles on the Gaia CMD in Figure~\ref{fig:cmd_age_wk2sdss}. In general, these objects follow the field sequence, including a large number of late Ms (M6, M7, M8). The color range of ($G - G_{\rm RP}$) = 1.4$-$1.6  is poorly sampled in other surveys outside of the youngest associations such as $\rho$ Ophiuchus and Upper Scorpius. The late Ms from K2SDSS are all within 200 pc, and their positions on the CMD and that of the other objects in the K2SDSS sample are consistent with the sample comprising unassociated field stars. 

\subsection{Rotation Period and the Gaia Color-Magnitude Diagram}\label{sec:rotgaiacmd}

\begin{figure}
    \centering
    \includegraphics[width=\textwidth]{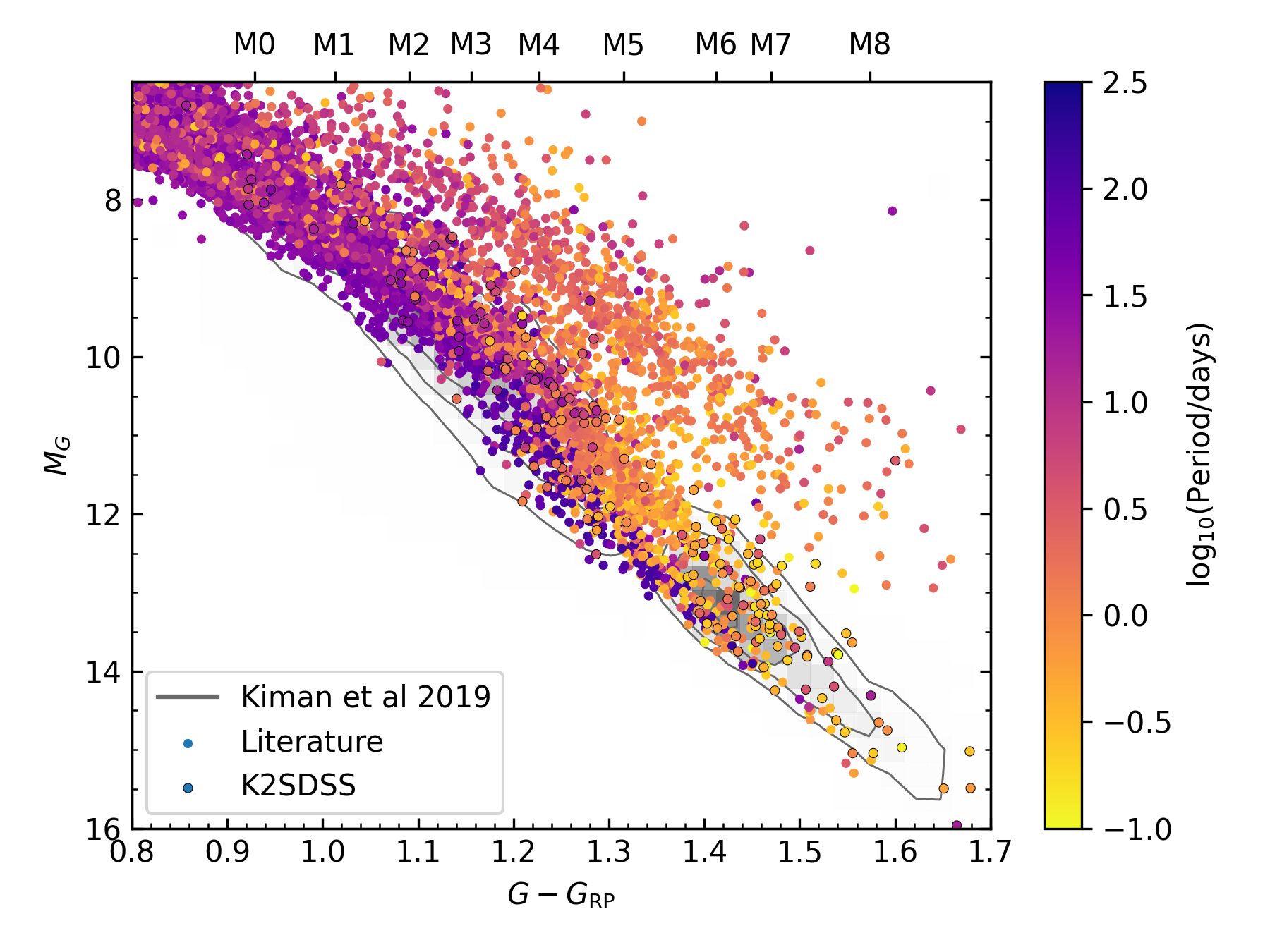}
    \caption{Color-Magnitude Diagram color coded by rotation period (yellow for fast rotators and purple for slower). Rotation rates are generally mixed, which is unsprising given the range of ages in the diagram. There is an overdensity of fast rotators at $G - G_{\rm RP}$ = 1.2--1.4}
    \label{fig:cmd_rotation_w}
\end{figure}

Figure \ref{fig:cmd_rotation_w} presents a Gaia DR2 CMD with our K2SDSS sample color-coded by rotation periods. Slowly rotating objects (purple) exist alongside neighboring stars with fast rotations (colored yellow) on the main sequence, across almost all ranges of colors and magnitudes. This contrasts with Figure \ref{fig:cmd_age_wk2sdss} where the CMD is colored by age and reveals layers of stellar sequences across color and magnitude. This is because rotation period is not tightly correlated with age for young M dwarfs, and CMD-position is not tightly correlated with age for old M dwarfs. This color and age dependence for the rotation period of stars (eg. \citealt{vansader2019}, \citealt{angus2015}) is difficult to see at a glance from Figure~\ref{fig:cmd_rotation_w} alone. A thorough description of this relation along with color-rotation plots is presented in Section \ref{6rotation_age}.

The largest concentration of fast rotating objects ($P_{\rm rot} < 2$ days) in Figure~\ref{fig:cmd_rotation_w} is at ($G - G_{\rm RP}$) = 1.2--1.4, between $M_G$ = 12.5--10.5. These objects are almost exclusively members of young associations (ages $<$ 700 Myr). Additionally, the fainter $M_G$ bound of the main sequence to ($G - G_{\rm RP}$) = 1.4 consists of mostly slow rotating objects in the sample. Figure~\ref{fig:cmd_age_wk2sdss} shows that the objects are most likely field sources with an age of at least a few Gyr. In the following section, samples are deliberately divided by age to explore their rotation period distribution.

%%%%%%%%%%%%%%%%%%%%%%%%%%%%%%%%%%%%%%%%%%%%%%%%%%%%%%%%%%%%%%%%%%%%%%%%%%%%%%%%%%%%%%%%%%%%%%%%%%%%%%%%%%%%%%%%%%%%%%%%
\section{M Dwarf Rotation Rates Versus Age}\label{6rotation_age}

The M dwarf rotation rates that are collected in this sample have all been analyzed in their original publications. Using Gaia colors for an analysis of every object is novel to this work, specifically the ($G - G_{\rm RP}$) color that has the tightest relation to spectral type for M dwarfs \citep{kiman_2019}. Previous works have looked at comparisons between clusters of different ages (e.g., \citealt{rebull2018usco,douglas_2019}), and field samples (ex. \citealt{kado-fong16}).

This paper treats clusters of known ages as independent snapshots in time, assuming them to be representative of M~dwarf rotational evolution at a given age. Differences due to metallicity, formation history, and interactions with other associations could all be ways to affect the angular momentum evolution and could impact the rotation period distribution of a given cluster, but evaluating those influences are beyond the scope of this work and will require a significantly larger ensemble of clusters. The goal of the following analysis is not to carryout a comparison of clusters, but to describe broad trends across the M dwarf rotational history by investigating samples of well-defined ages.

\begin{figure}
    \centering
    \includegraphics[width=4in]{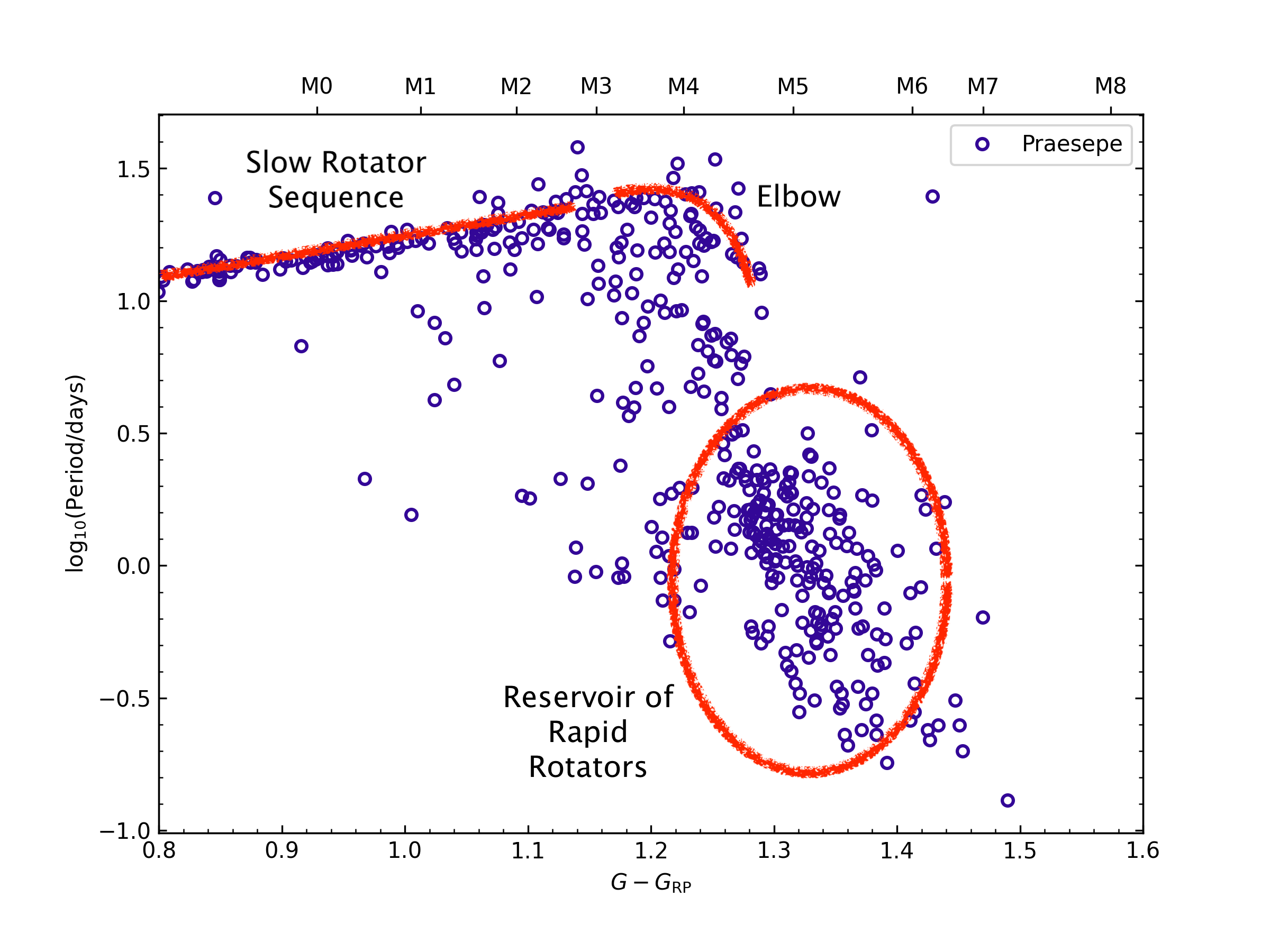}
    \caption{Rotation period distribution for Praesepe from \citet{douglas2017}, with additional labels for the terms to be used to discuss rotation period distributions in the following sections.}
    \label{fig:praesepe}
\end{figure}

For context, Figure \ref{fig:praesepe} presents the rotation rates of the Praesepe cluster from \citet{douglas2017} against Gaia DR2 ($G - G_{\rm RP}$) color. There are three major features that can appear in the rotation period distributions of the low-mass star population in a given cluster, the slow rotator sequence, the elbow, and the reservoir of fast rotators.

In a group of coeval stars, the sequence of slowly rotating stars (the $I$ sequence according to \citealt{barnes_2003}) are objects that have converged to a similar period compared to other objects of the same mass and color. While the convergence of stars to a similar period is the defining feature of the slowly-rotating sequence, slow is a relative term and the rotation period is age- and mass-dependent. For example, F-type stars in Praesepe have converged to periods of just a few days, which is far faster than early Ms which converged at $15-20$ days. In Figure~\ref{fig:praesepe}, the $\grp$ axis serves as a mass proxy, and the slowly rotating sequence begins from ($G - G_{\rm RP}) \approx 0.8$ and extends through to ($G - G_{\rm RP}) \approx 1.25$ ($\approx$M4 spectral type). While Figure~\ref{fig:praesepe} focuses on M dwarfs, the slow-rotator sequence stretches into the bluer and more massive F, G, and K stars \citep{douglas_2019}.

\begin{figure}
    \centering
    \includegraphics[width=\textwidth]{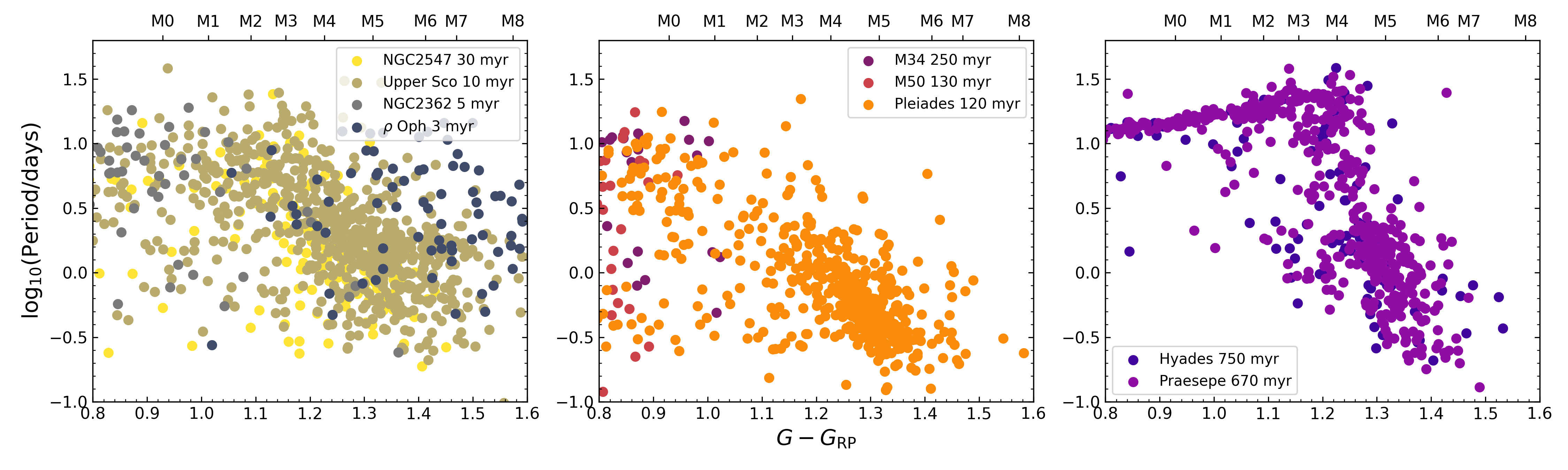}
    \caption{A series of color--rotation period distribution plots for the cluster samples, separated into < 100 Myr, 100Myr to 500 Myr, and $>$ 500 Myr groups. While all have rapid rotators in $G - G_{\rm RP}$ = 1.2-1.4, the evolution of the elbow or at what color objects begin to converge onto the slow rotator sequence, moves redwards with age. }
    \label{fig:series}
\end{figure}

\begin{figure}
    \centering
    \includegraphics[width=5in]{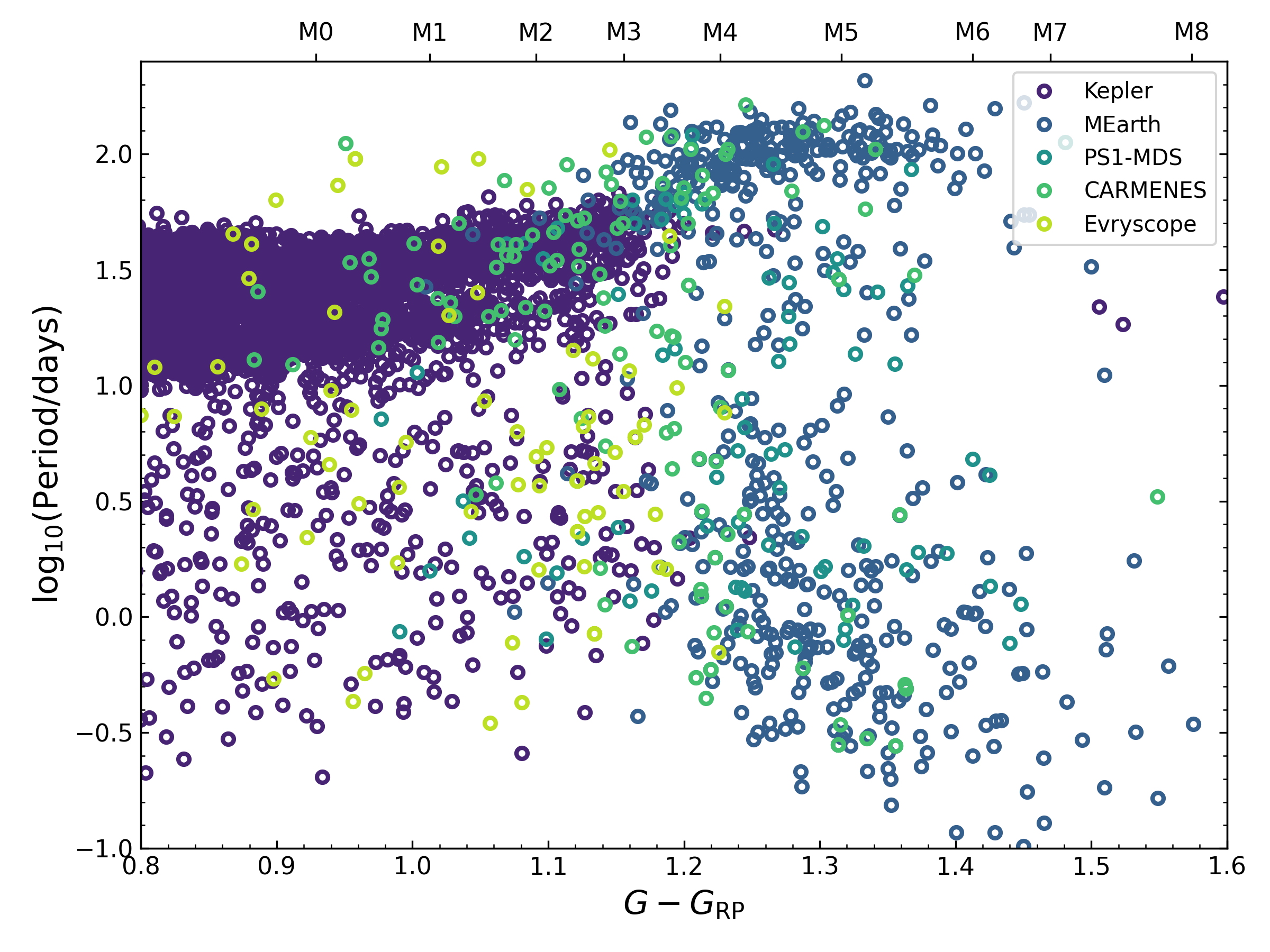}
    \caption{Color - rotation period distribution for the field portion of the sample. Each literature is plotted in a distinct color. The larger number of Kepler objects dominate for the early Ms, but bimodality in period of the later Ms first noticed in \citet{newton2016} is recovered in all samples.}
    \label{fig:field_rot_color}
\end{figure}

A critical feature in the period distribution of Praesepe members occurs at $G - G_{\rm RP} \approx 1.25$, where redder stars are no longer tightly converged, or even converged at all. We define the elbow as the color where objects are beginning to converge onto the slow rotator sequence, even if there is a large spread of rotation rates at that color. While it is relatively clear in Figure~\ref{fig:praesepe}, for the period distributions that follow (Figure~\ref{fig:series}-~\ref{fig:color_per_log_n}), the elbow can be less defined, and even absent for the youngest clusters. 

The final feature that is common in many clusters, labeled as the "reservoir" in this work, is defined for Praesepe by the rapidly rotating ($P_{\rm rot}$ $\leq$ 2 days) objects in the ($G - G_{\rm RP}$) color range between $1.2-1.4$. It is the largest area of age degeneracy in color-rotation space across our entire sample, with young cluster objects at only 10 Myr displaying similar rotation periods to field-aged stars of the same color. The reservoir of rapid rotators discussed in this work is not necessarily part of the \textit{C} sequence discussed in \citet{barnes_2003}, because while they rotate faster than objects on the slow rotator sequence, their rotation rates are not converged and vary from $P_{\rm rot}$ $\sim$0.3 -- 2 days.

In Figure~\ref{fig:series}, clusters of similar ages are shown together. The first panel features clusters with ages $leq$ 30 Myr (NGC2547, Upper Scorpius, NGC2362, $\rho$ Ophiuchus), the second panel has clusters with ages between 100 Myr and 250 Myr (M34, M50, Pleiades) and the final panel has clusters with ages $>$ 500 Myr, (Hyades, Praesepe). 

Clusters with ages younger than 100 Myr are in the first panel of Figure \ref{fig:series}. These objects are still converging onto the main sequence. There is no evidence of convergence in rotation rates at any color in this range of ages, and thus there is no slow rotator sequence or elbow in any of these clusters. \citet{rebull2018usco} also noted a trend in $P_{\rm rot}$ vs ($V$-$K_s$) color for Upper Scorpius where redder objects have faster rotation rates. This trend is also seen in our period vs ($G - G_{\rm RP}$) diagram for Upper Scorpius, as well as NGC 2362 and NGC 2547, especially at $\grp \approx 1.2$. Overall, there is far less structure in this range of ages compared with older clusters, as our sample shows a wider range of rotation rates for a given color. Stars likely form with a wide array of angular momenta \citep{herbst_07,lamm_2007_onc_ngc2264}, and this range of angular momenta is likely driving the spread in color--period distribution at this young age. Another feature at this age is a pile up of rotation rates around log $P_{\rm rot}$ = 0.3 in Upper Scorpius for objects with infra-red (IR) excess first found in \citet{rebull2018usco}. This was interpreted as magnetic braking between the circumstellar disk of the young star with its magnetic field. In the same work the objects with IR excess in the younger $\rho$ Ophiuchus did not show a difference in their distribution. Looking at the rotation rate distributions of NGC 2547 and NGC2362 in Figure ~\ref{fig:series} we do not find any overdensity in rotation, but those clusters have far fewer objects than Upper Scorpius both in total number of sources and in number of objects reported with IR excess. The magnetic locking of a disk will matter for the angular momentum evolution of an individual star and it will likely imprint a signature on the gyrochronology relation as seen by \citet{rebull2018usco} in Upper Scorpius.  However it is unclear how that signature evolves across age or how it presents itself in known clusters of different ages. 

The second panel of Figure \ref{fig:series} contains clusters with ages between 100 Myr and 500 Myr. At this age M dwarfs should all be on the main sequence, with only minor contamination from substellar mass objects, as described in Section~\ref{sec:sample}. It is difficult to describe their respective elbows because the slow rotator sequence is not as well-constrained in any of the associations, with the early Ms still having rapid rotators. While there is no sharp transition between converged objects and quick rotators in any of the clusters, a fraction of stars have similar rotation rates compared with converged stars up to ($G - G_{\rm RP}$)= 0.9 in the Pleiades, M50, and M34. The Pleiades has many objects in its rapid rotating reservoir, while M34 an M50 do not have members at that color. This is due to the mid- to late-Ms in M34 and M50 not satisfying our quality cuts as they are significantly further away. The number of $\approx$ 5 day rotators decreases in the Pleiades after M2 when compared to the first panel. \citet{rebull2018usco} previously found a trend in Pleiades members where an increase in rotation rate was observed at redder ($V$-$K_s$) colors. We see an agreement with that result, as objects at ($G - G_{\rm RP}$) = 1.2 are spinning slower than those at 1.4.

The final panel has the oldest objects, with the slow rotator sequence, elbow and reservoir present in Praesepe and Hyades. The slow rotator sequence is at a longer period for an older cluster at a given ($G - G_{\rm RP}$) , and it has been shown \citep[e.g.,][]{curtis2019,douglas_2019}in the FGK mass range to serve as a successful relation between rotation and age. The elbow in these clusters is at $\grp \simeq 1.25$ for Praesepe, and 1.28 for the Hyades. For M34 it is difficult to identify a well-defined elbow, but there may be convergence up to ($G - G_{\rm RP}$) = 1.0 just at the start of the M dwarf regime. The elbow appears to move redwards in $G - G_{\rm RP}$ with older ages. Praesepe and Hyades both have M dwarfs in the rapid rotator reservoir. 

The rotation period distribution of the field samples is presented in Figure~\ref{fig:field_rot_color}. The largest fraction originates from the Kepler sample from \citet{mcq2014}. While field objects have a range of ages, their average age is older than our cluster samples. There is a noticeable lack of fast rotators at log $P_{\rm rot}$ $\geq$ 1.0 day and $\grp \gtrapprox 0.8$, that becomes even more apparent with redder colors. This "edge" in the field sample could be caused by a slow rotator sequence. For example, in Figure~\ref{fig:praesepe} where Praesepe's rotation period distribution is shown there are very few quickly rotating objects (log $P_{\rm rot}$ $<$ 1.0) at colors where objects have converged onto the slow rotator sequence ($\grp = 0.8-1.1$). As the objects get older, they are expected to continue losing angular momentum and slowing down due to magnetic braking. The edge of the Kepler sample may be the fastest bound of the slow rotator sequence, with older stars evolving onto the slow rotator sequence and filling out the space at slower rotation periods. While the sharp edge of the Kepler field objects ends around ($G - G_{\rm RP}$) = 1.18 due to the selection function of the survey, the other field samples seem to suggest it continues to redder colors by the longer period MEarth objects. \citep{newton2016} 

Bimodality in the rotation rates \citep{newton2016,kado-fong16,evryscope_2020} is another feature of the late Ms. In the ($G - G_{\rm RP}$) $>$ 1.2 color range, most objects are either rapid rotators ($P_{\rm rot}$ $<$ 2 days) or slowly rotating ($>$ 100 days). \citet{newton2016} explains this in terms of a relatively sudden jump of late M dwarfs onto the slow rotator sequence. This is therefore suggestive of a quick transition from rapid to slowly rotating M dwarfs. The growing sharpness of the elbows in older clusters also points towards faster transitions from fast to slow rotators in this regime. We discuss this feature and how it compares to theoretical work \citep[e.g.,][]{matt_2011,matt_2015,garraffo_2018_revrev} in Section \ref{sec:standardpic}. This transition must happen after $\approx$700 Myr because our oldest cluster contains stars that are still fast rotators at ($G - G_{\rm RP}$) $>$ 1.25 and no evidence for $>$ 100 day periods at that color. The evidence and potential causes for this transition is explored further in Section \ref{sec:standardpic}.

\subsection{The SDSS Sample in context with Cluster and Field Stars}

\begin{figure}
    \centering
    \includegraphics[width=7in]{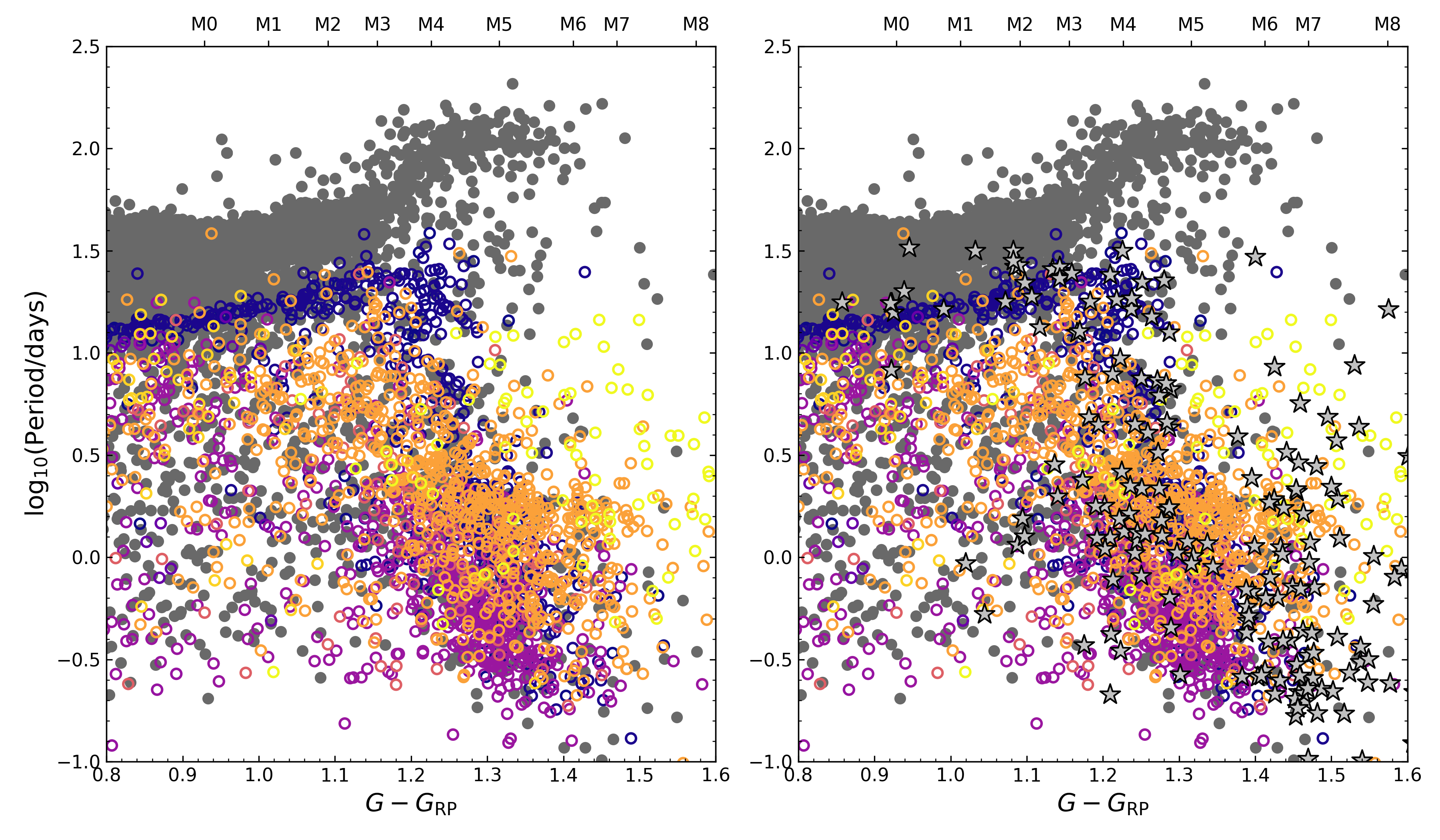}
    \caption{Color--period distribution of our full literature sample on a logarithmic scale, then with K2SDSS overlaid. Associations with known ages are colored youngest (yellow) to oldest (purple), with field stars in grey.}
    \label{fig:color_per_log_n}
\end{figure}

Figure~\ref{fig:color_per_log_n} shows all stars of our sample in a color-rotation diagram including the K2SDSS sample. As discussed in Section \ref{sec:sample}, 26 objects in the K2SDSS sample are known members of Praesepe, most recently analyzed in \citet{douglas2014} and \citet{douglas2017}. Comparison between previously measured rotation rates and the rotation periods measured in this work  agree on average within 3.4\%. The rest of the K2SDSS stars are assumed to be field stars.

There are several early M dwarfs in the K2SDSS sample that have rotation rates slower than that of Praesepe members at every given color in Figure \ref{fig:color_per_log_n}. Since stars of the same color have converged onto the slow rotator sequence in Praesepe, it is straightforward to assume that these stars that are rotating more slowly have converged as well and are therefore older, but we do not attempt to quantify how much older. Between ($G - G_{\rm RP}$) 1.1-1.4 in Figure \ref{fig:color_per_log_n} there are K2SDSS objects at long rotation periods of 20-30 days similar to objects in the elbows of 600 Myr clusters, as well as many objects with $P_{\rm rot}$ $<$ 2 days in the rapid rotator reservoir. 

The large number of late M dwarfs (M6,M7,M8) that are unique to the K2SDSS sample (see section \ref{sec:rotation_k2sdss}) are mostly rapid rotators, with $P_{\rm rot}$ $<$ 2 days. This is an undersampled region compared to the rest of the color space in the M~dwarf regime, and the quick rotation rate seems to extend the range of the reservoir. They add to the bimodality seen in field samples mentioned in the previous subsection.

%%%%%%%%%%%%%%%%%%%%%%%%%%%%%%%%%%%%%%%%%%%%%%%%%%%%%%%%%%%%%%%%%%%%%%%%%%%%%%%%%%%%%%%%%%%%%%%%%%%%%%%%%%%%%%%%%%%%%%%%
\section{Rotation periods, H$\alpha$ and Kinematics}\label{7halphakin}

\subsection{Chromospheric H$\alpha$ emission}

\begin{figure}
    \centering
    \includegraphics[width=6in]{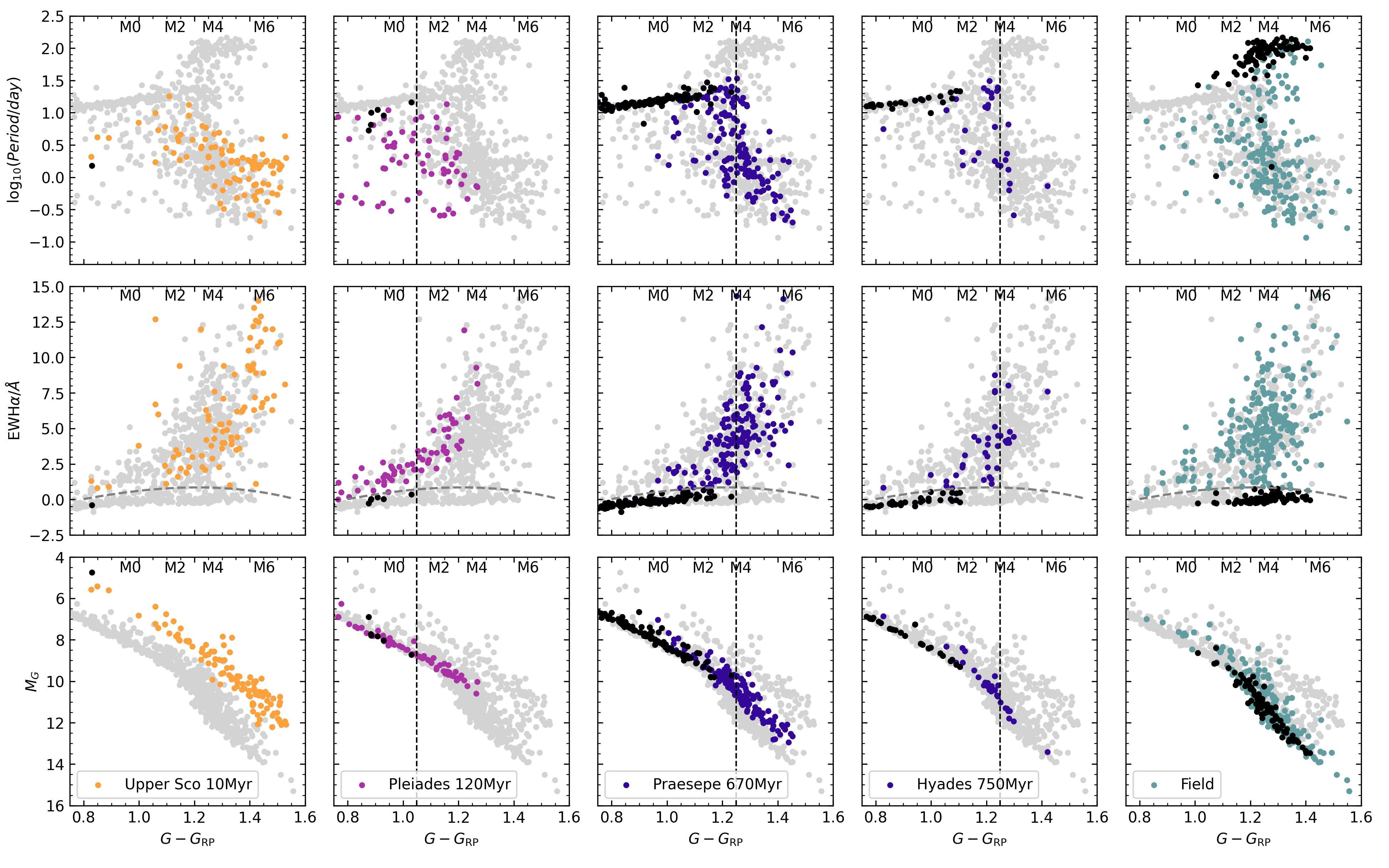}
    \caption{Grid with Upper Scorpius, Pleiades, Praesepe, Hyades, and field objects with log rotation period, H$\alpha$ equivalent width and $M_{G}$ magnitude versus $G - G_{\rm RP}$. The H$\alpha$ activity boundary \citep{kiman_2019} is included in the second row, with objects considered inactive plotted as black points in every panel. Clusters are presented in increasing age order from left to right, with the field objects likely being a mix of ages but also including the oldest.}
    \label{fig:grid_plot}
\end{figure}

Chromospheric emission is also correlated with age because magnetic activity is driven by rotation; as stars spin down over time, their chromospheric emission decays \citep[e.g.,][]{skumanich_1972, mamajek_hillenbrand}. 
For M dwarfs, H$\alpha$ emission is the standard indicator of chromospheric activity \citep{West_2008, west_2015_activity_rotation, kiman_2021}. \citet{eggen_1990_halpham} found H$\alpha$ decreased with age,  while \citet{west_2006_m7magnetic} found the fraction of active H$\alpha$ stars decreased with galactic vertical height. \citet{,newton_2017_halpha} found a power law decay in fractional H$\alpha$ luminosity ($L_{\rm H\alpha}/L_{\rm Bol}$) in M dwarfs, and showed that a non-detection of H$\alpha$ in a given M dwarf is correlated with a slow rotation. 

Figure~\ref{fig:grid_plot} shows H$\alpha$ equivalent width (we take the values compiled in \citet{kiman_2021} for Pleiades, Upper Scorpius, Hyades and multiple field objects, as well as \citet{douglas2014} for Praesepe and \citet{newton_2017_halpha} for MEarth objects in the field), rotation period, and $M_{\rm G}$, versus ($G - G_{\rm RP}$) color. Sources that have measurements for all variables are in grey, and each column is a subset of our sample highlighted in an arbitrary color.

Objects are defined as inactive if the H$\alpha$ equivalent width is below the activity relation from \citet{kiman_2021}, which is based on fitting a third-order polynomial to SDSS data. We do not differentiate between various instruments in regards to the H$\alpha$ measurements. For the purposes of our analysis this is appropriate---redefining  our inactivity boundary to other common levels (e.g., $0.5-1.5$ \AA) would only result in a handful of objects for each sample being categorized differently, and therefore does not impact our main conclusions.

As in \citet{kiman_2021}, the most active objects in clusters are those with the latest spectral types; the H$\alpha$ EW decreases towards bluer colors, and therefore with the more massive M dwarfs. For the clusters presented here, we have included a vertical dashed line at the approximate color at which the rotation period elbow occurs (see Section \ref{6rotation_age}; e.g., $\grp \approx 1.05$ for Pleiades, $\grp \approx 1.25$ for Praesepe). The elbow approximately aligns with where objects transition from inactive to active in their H$\alpha$ equivalent width, and the line drawn from it marches redward down the main sequence of the CMD.

Inactive stars in the H$\alpha$ versus color plot (middle panel of Figure~\ref{fig:grid_plot}) lie on the slow rotator sequence in the clusters in the rotation versus color plot (top panel of Figure~\ref{fig:grid_plot}). In addition, while there is not a well defined sequence for the field sample, all the inactive stars have rotation periods far larger than the rapid rotators ($P_{\rm rot}$ $<$ 2 days). The converse is also true: there are no active stars that are on the slow rotator sequence, or with a rotation rate that would place them in the rapid rotator regime, in agreement with \citet{newton_2017_halpha}.

Chromospheric activity is driven by the magnetic field of the star, but can also be impacted through interactions with a close companion. The dynamo that drives the stellar magnetic fields depends on the rotation of the star \citep[e.g.,][]{charbon_dynamo}, and so any phenomenon that could change the rotation rate will also affect its magnetic activity. \citet{douglas2017} found outlying rotation rates in Hyades to mostly be from candidate or confirmed binaries, while \citet{stauffer_bin_18} found evidence that binary members of Upper Scorpius and Pleiades spin faster than average members, although by the age of Praesepe the significance was reduced. Therefore, for young stars rotation rate and chromospheric activity could be increased for M dwarfs in binary systems.

\subsection{Galactic Kinematics}

We examine objects from the full sample of rotating M dwarfs using their Galactic kinematics and their H$\alpha$ equivalent width measurement where available. The 3D velocity of a population of stars can serve as an age indicator, as galactic orbits become kinematically heated over time due to gravitational interactions with other stars and giant molecular clouds \citep{aumer_binney_solarkine, ting_rix_disk, angus_2020}. 

While DR2 does not have radial velocities for every object, which prevents full 3D velocity calculations, all of the objects in the sample have parallaxes and proper motions which enables a calculation of tangential velocity ($v_{\rm tan}$) for all of the objects in the rotation sample, using the following equation:

\begin{equation}\label{eq:kin}
   v_{\rm tan}  = 4.74\;d\;\sqrt{\mu_{\rm R.A.}^2 + \mu_{\rm decl.}^2}
\end{equation}

\noindent where $v_{\rm tan}$ has units of \kms, the proper motions ($\mu$) have units of \mas, and the distance ($d$) has units of pc and is calculated from the inverse parallax.

\citet{angus_2020} has shown that kinematics can be used to trace gyrochrones across the Kepler \prot\ distribution \citep{mcq2014}. Also, \citet{faherty09} and \citet{kiman_2019} each used kinematics to probe the ages of their samples, late-Ms and brown dwarfs and SDSS M dwarfs respectively. Here, we examine how kinematics indicate age characteristics of the whole M dwarf sample.

\subsubsection{Cluster Kinematics}

\begin{figure}
    \centering
    \includegraphics[width=\textwidth]{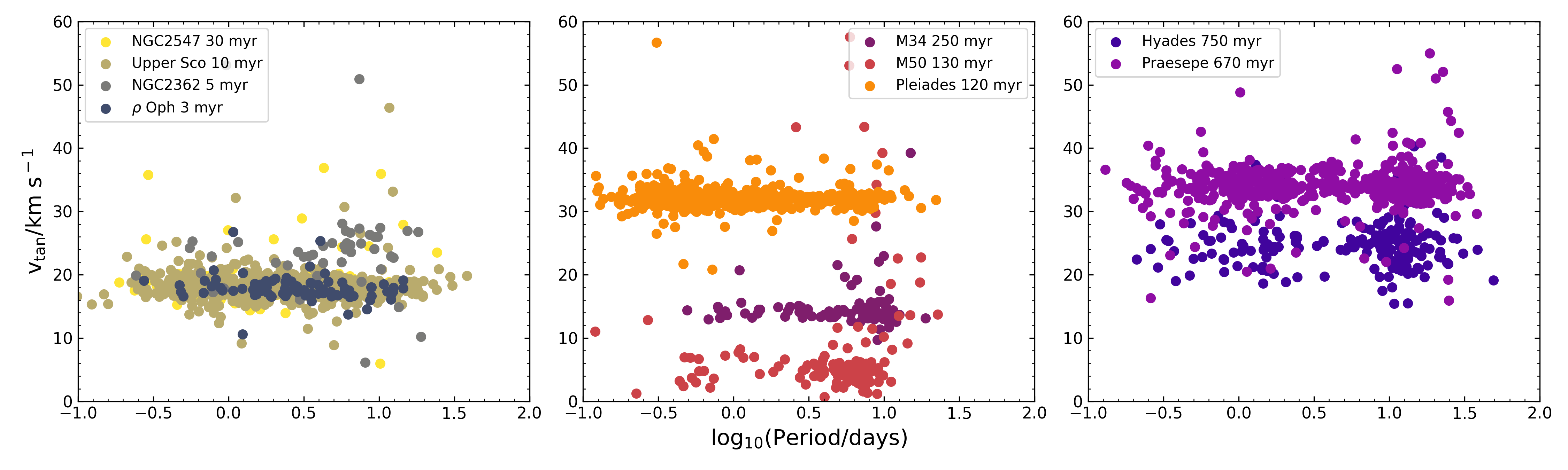}
    \caption{Tangential velocity and rotation period distribution of the cluster M dwarfs sample. Low variance in velocities is used to define cluster memberships, and our cluster objects reflect this with extremely similar tangential velocity to other members.}
    \label{fig:clusterkinematics}
\end{figure}

In Figure~\ref{fig:clusterkinematics}, we present the $v_{\rm tan}$ values of the clusters in the sample as a function of $\log \prot$. As cluster membership is often defined in part by similarities in kinematics, each cluster shows only a small dispersion around a respective mean value, with an average standard deviation within a cluster of 5~\kms.\footnote{This is significantly larger than their internal velocity dispersions ($\sigma_V \lesssim 1~\kms)$, and is likely caused by the inclusion of binaries and stars with poor astrometric solutions in the calculations.}  
The objects that are outliers from a clusters mean velocity are likely contamination either in the original membership list or due to our cross match. The $v_{\rm tan}$ and its dispersion are not useful as diagnostics of relative age between clusters, or absolute ages of populations of stars but are all consistent with kinematically young objects with an average $v_{\rm tan} < 40$~\kms.

\subsubsection{Field Sample Kinematics}

\begin{figure}
    \centering
    \includegraphics[width=6in]{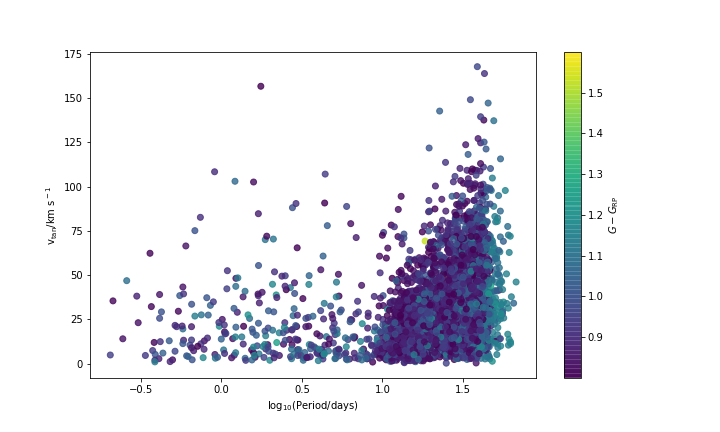}
    \caption{Tangential velocity and rotation period distribution of the Kepler M dwarf sample. Objects with rotation rates below 10 days are poorly sampled compared to the rest of the sample, which shows an increase in dispersion of tangential velocity with longer rotation period.}
    \label{fig:keplerkinematics}
\end{figure}

We considered the field sample in two parts: 1) the Kepler sample, and 2) non-Kepler sample. The Kepler objects outnumber other field objects in the sample approximately three to one, therefore we decided to consider them separately. We plot $v_{\rm tan}$ against $\log P_{\rm rot}$ for the Kepler objects in Figure~\ref{fig:keplerkinematics}. We note a trend in the increase in the median and the dispersion of $v_{\rm tan}$ with $\log P_{\rm rot}$ in the Kepler field, especially after $\log P_{\rm rot}$ of 1.0, or 10 days. A rotation period of 10 days is important in the Kepler M dwarfs, as the vast majority of the sample is greater than 10 days. Ten days is also the rotation rate of late K/early M objects in Praesepe, which are already on the slow rotator sequence. Indeed, the Kepler sample extends only to a ($G - G_{\rm RP}$) value of 1.2, which is before the elbow in Praesepe. The Kepler sample with rotation rates longer than 10 days likely consists of only converged stars on their respective slow rotator sequences, and it is within this $>$ 10 day regime that we focus our analysis. 

\begin{figure}
    \centering
    \includegraphics[width=\textwidth]{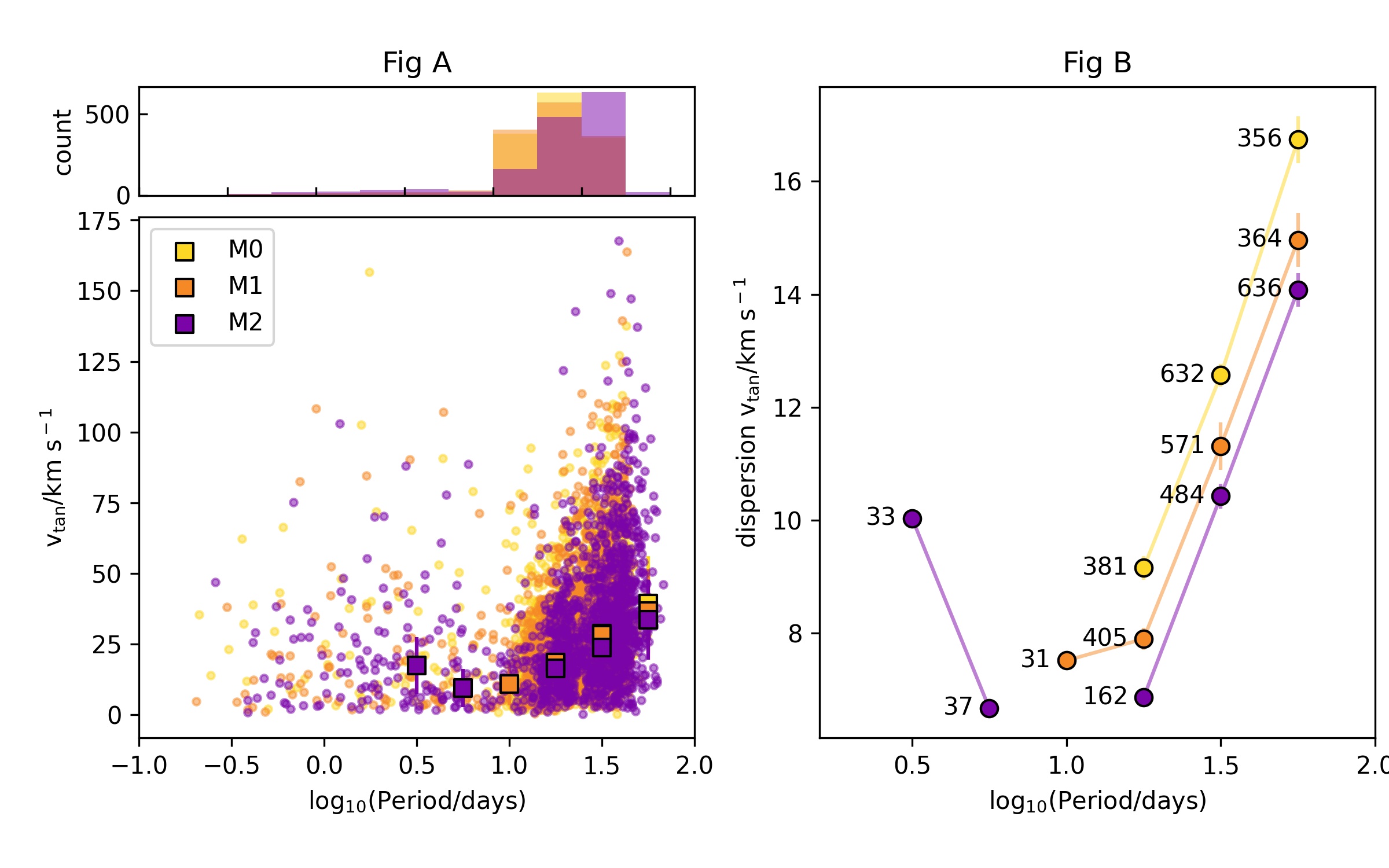}
    \caption{(\textit{left}) Tangential velocity and log rotation period distribution of the Kepler M dwarf sample binned by $G - G_{\rm RP}$ color associated with M0, M1, and M2 spectral types. Median $v_{\rm tan}$ per 0.5 $\log \prot$ bins is overplotted. We only include bins with over 30 objects. 
    (\textit{right}) Median average deviation as dispersion for the same bins.  For the well sampled rotation rates above 10 days later spectral bins show less $v_{\rm tan}$ dispersion at the same rotation rate compared to earlier spectral type bins.}
    \label{fig:kepler_bin_kinematics}
\end{figure}

In Figure \ref{fig:kepler_bin_kinematics}, the left panel are M dwarfs in the Kepler field binned by their \grp\ color corresponding approximately to M0, M1, and M2 bins, with splits at \grp\ = 0.87 and 0.98. The right panel shows the median absolute deviation as the dispersion for the median $v_{\rm tan}$ values. Errors on the dispersion were calculated through a bootstrap method, sampling with replacement 100 times. We do not calculate dispersion for bins with fewer than 30 objects. Each bin with $\log \prot$ $>$ 1 shows an increase in both median $v_{\rm tan}$, and $v_{\rm tan}$ dispersion with longer rotation period. Additionally, for a given $v_{\rm tan}$ dispersion in one bin, the same value occurs at longer rotation periods in a redder color bin. This agrees with \citet{angus_2020}, which showed the same result with galactic latitude velocity dispersion compared to effective temperature and mass for higher mass stars. This is caused by the fact that once objects are converged onto the slow rotator sequence in the early M regime (Praesepe age and older for M0-M2) lower-mass stars rotate more slowly at a given age than higher-mass stars.

\begin{figure}
    \centering
    \includegraphics[width=6in]{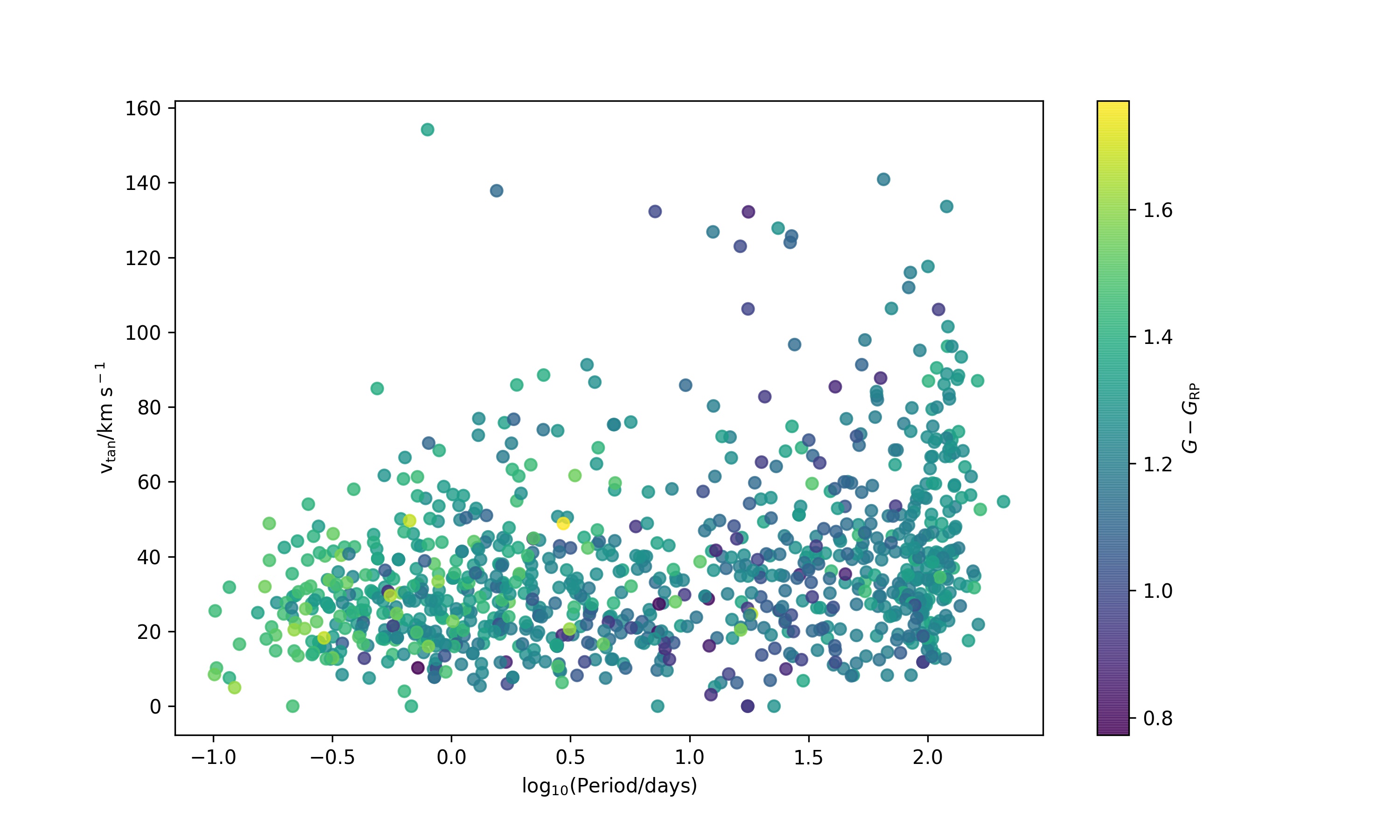}
    \caption{The rest of the field tangential velocity and rotation period period distribution. While there are far fewer objects they also cover longer rotation periods compared to the Kepler sample}
    \label{fig:field_kin}
\end{figure}

The rest of the field sample is plotted with $v_{\rm tan}$ against $\log \prot$ in Figure \ref{fig:field_kin}. K2SDSS does not have the baseline to be sensitive to greater than ~45 day periods, and we do not attempt to explore or correct for the selection functions of any of the samples. Overall the $v_{\rm tan}$ dispersion increases with $\log \prot$, which is evidence for kinematic heating with age for the population of longer rotation objects. 

\begin{figure}
    \centering
    \includegraphics[width=6in]{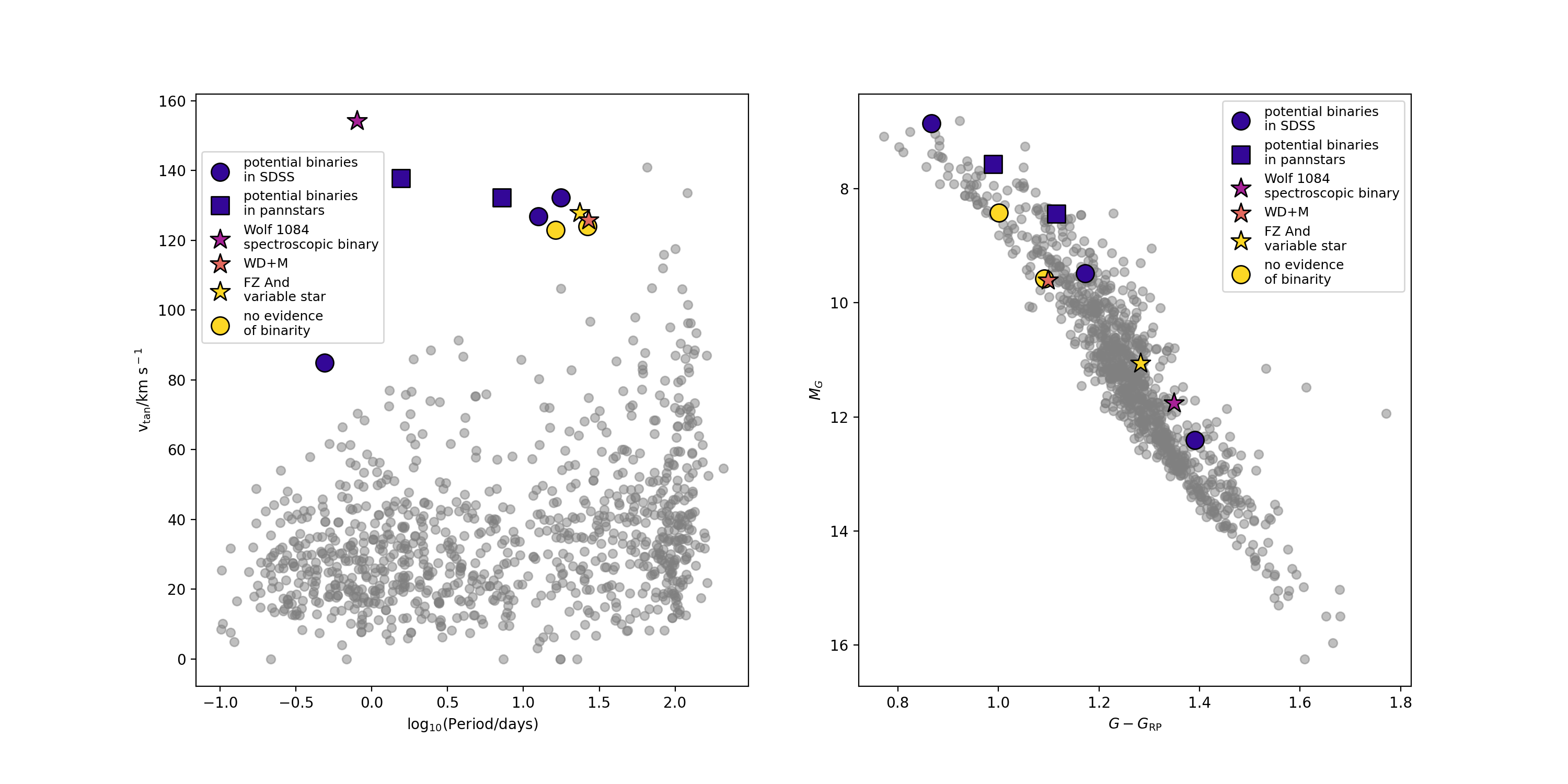}
    \caption{Highlighting outliers in the field $v_{\rm tan}$ - $\log \prot$, and where they fall on the CMD compared to the rest of the field sample.}
    \label{fig:field_kin_binary}
\end{figure}

There are several exceptions that have large $v_{\rm tan}$ values but faster than expected rotation periods that are colored in Figure \ref{fig:field_kin_binary}. All of the objects appear as single sources in Gaia DR2. The purple five point star is Gaia DR2 2183807118442110720, a spectroscopic binary (Wolf 1084). Gaia DR2 1992715605301433856 (FZ And) is an eruptive variable star in yellow plotted as a five point star. Gaia DR2 663264446538001792 is a white dwarf and M dwarf spectroscopic binary system plotted as an orange five pointed star \citep{reb-man_wd}. The yellow circles (Gaia DR2 2536789589768291968 and 600909600234425472) and blue circles (Gaia DR2 608202897083854592, 3794059833092343168, and 2183807118442110720) are all from the K2SDSS sample and have photometric distances from SDSS, and parallax distances from Gaia DR2. The gold objects have consistent distances from the two methods within 2 sigma, while the red objects do not. \citet{paudel2019} used the inconsistency between the two distance measurements as evidence for binarity in the binary system EPIC 248624299. However they also saw a beat period in the light curve, and none of the yellow or blue objects from our sample show evidence of a second rotational signal. Finally Gaia DR2 1429320143507716096 and 855574566449072256 have so far only been referenced in \citet{kado-fong16}, the same source as their rotation rates, and are plotted as blue squares. They appear as isolated sources in Gaia DR2, but both have a relatively low short peak ratio, a metric used in \citep{kado-fong16} to distinguish the strongest peak of the periodogram. If they are binary systems with an unresolved companion with detectable rotation variability, they would have low short peak ratio as well (see Section \ref{sec:binaries} for examples of ratio of Lomb--Scargle power in binaries for the sample in this work). \citet{kado-fong16} estimated a 15\% contamination of unresolved binaries in their sample, but did not attempt to identify probable binaries. We plot the field CMD in Figure \ref{fig:field_kin_binary} with the objects listed similarly colored. The field sample does not have a tightly constrained main sequence, nor a clear binary track. However, all the objects colored blue are amongst the most luminous for their ($G - G_{\rm RP}$) colors, which could elevate them to candidate binaries, as is the yellow 5 point star FZ And. Binary systems are known to effect the rotation of the members (\citet{stauffer_bin_18} for example), and could explain why they appear kinematically old while still having relatively quick and potentially young rotation rates.

\begin{figure}
    \centering
    \includegraphics[width=6in]{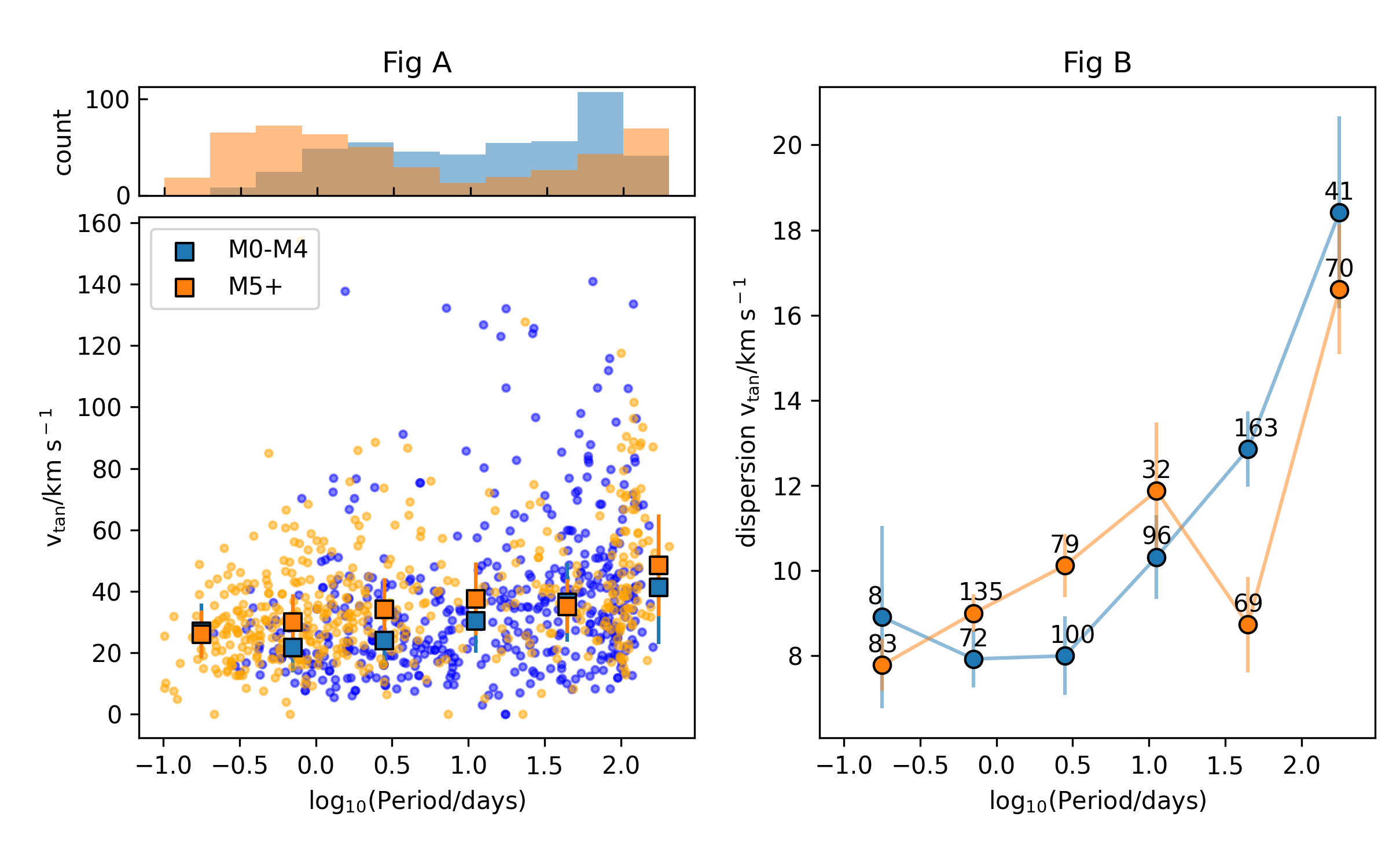}
    \caption{Same as \ref{fig:kepler_bin_kinematics} but for field M dwarfs with two $G - G_{\rm RP}$ color bins split at $G - G_{\rm RP}$ = 1.26 representing early and mid M dwarfs. The greater dispersion at faster rotation rates in later M dwarfs (orange points) could be evidence for stalling at quicker rotation rates than for early Ms.}
    \label{fig:field_bin_kinematics}
\end{figure}

We further analyze the field sample by dividing it into two bins: M dwarf objects with ($G - G_{\rm RP}$) colors of 1) $\leq$ 1.27 and 2) $>$ 1.27. We plot them with $v_{\rm tan}$ and $\log \prot$, along with $v_{\rm tan}$ dispersion in Figure \ref{fig:field_bin_kinematics}. This division was chosen based on the \citet{kiman_2019} mean spectral type colors for M dwarfs to roughly divide the objects into M0-M4 and M5+ spectral types. This is also near the partially/fully convective boundary \citep{charb_baraffe}. The distribution in $\log \prot$ space of the two bins is also plotted, and both populations show the general bimodality in M dwarf rotation rates first noticed in \citet{newton2016}. The faster rotating peak is at $\log \prot$ of 0.25 or a period of 1.7 days for the earlier types compared to the $\log \prot$ of 0.33 or a period of 0.46 days for the later types, with both bins having the longer peak in the bimodal distribution at $>$ 100 days. Despite their bimodality, the earlier spectral types are more evenly distributed across $\log \prot$ space, compared to the later types. The bimodality for M dwarf objects is apparent across each literature source included in this sample (see Table \ref{source}), and was postulated in \citet{newton2016} to be due to a stalling of rotational evolution in M dwarfs before a rapid transition to longer rotations. 

We interpret the dispersion of $v_{\rm tan}$ for the two color bins as evidence for this stalling in late M dwarfs. At $\log _{10} P_{\rm rot}<1$  (or $<$ 10 days) the dispersion is consistently greater for the redder color bin than for the bluer. This area of rotation space is populated across the M dwarf regime in young clusters (see Figure \ref{fig:series}). While the tangential velocities are still mostly consistent with young ages, the larger dispersion for objects in the redder bin implies an older kinematic age than for objects in the bluer bin at that rotation rate. This could be due to bluer objects transitioning sooner to longer rotation periods as they converge to the slow rotator sequence, and redder objects stalling at shorter rotation rates and becoming kinematically heated over time. For rotation rates $>$ 10 days ($\log _{10} P_{\rm rot}>$ 1), we see increased $v_{\rm tan}$ dispersion with rotation rate consistent with kinematic heating over time for both bins, but similar to the Kepler sample, the bluer objects show greater dispersion for a given rotation rate. It is worth noting that even though the rotation bins were selected such that there were a statistically significant sample of objects within each bin, the number of objects in each rotation bin is relatively small. This is especially true for the bins between the peaks in the rotation period distributions of the later types, where one bin merely contains on the order of a dozen objects.

%%%%%%%%%%%%%%%%%%%%%%%%%%%%%%%%%%%%%%%%%%%%%%%%%%%%%%%%%%%%%%%%%%%%%%%%%%%%%%%%%%%%%%%%%%%%%%%%%%%%%%%%%%%%%%%%%%%%%%%%
\section{The standard M dwarf Gyrochronology picture}\label{sec:standardpic}

In this section, we synthesize the trends presented in the previous sections along with current theoretical approaches to understanding stellar spin down and present a summary of the standard M dwarf gyrochronology picture.

Gyrochronology is successful at predicting ages for stars such as F, G and K at young ages, as they quickly converge onto the slow rotator sequence and continue to slow down from there \citep[e.g.,][]{barnes_2003}. The time needed to converge onto the slow rotator sequence increases with decreasing mass, cooler temperature, and redder color;  depending on the age of the cluster, stars have only converged within a certain range of spectral types \citep[e.g.,][]{matt_2011,rebull2018usco,bouvier_2009}. As presented in Section \ref{6rotation_age}, this slow sequence does not begin to extend into the early M dwarfs until at least 200 Myr, 
as early M dwarfs in the Pleiades and M34 have not yet spun down and converged (see Figure \ref{fig:series}). 
Early Ms are not witnessed to have truly converged until $\approx$700 Myr (as seen in Praesepe and Hyades), 
but this is due to a large age gap in the benchmark cluster sample between 200-700 Myr.
This puts something of a lower bound on the application of M dwarf gyrochronology, where under 400-700 Myr for early to mid Ms, the distribution of rotation rates is dominated by a natal spread in angular momentum from formation \citep{kawaler1988,matt_2015}.

At ages of $>$700 Myr, we begin to see the split within the M dwarf regime, where earlier M dwarf spectral types converge and evolve and later ones are yet to do so. The color at which the M dwarfs have mostly converged appears as a sharp elbow: where bluer colors are converged, objects near the elbow appear to have intermediate rotation rates between fast rotators and the slow sequence, and redder colors consisting of later M dwarfs are still rotating rapidly in a broad distribution of rates. Because clusters provide snapshots in time we see these transitions. While traditional gyrochronology relations focus on the evolution of objects on the slow rotator sequence, there is no calibrated empirical gyrochronology relation for mid to late M dwarfs, as none have been observed to be converged on the slow rotator sequence in benchmark clusters. 

There is possibly a change in the distribution of rapid rotation rates in M dwarfs across clusters, presented in \citet{rebull2018usco}. However, a description of the morphology of the wide distribution of rapid rotation rates of an ensemble of similarly aged objects is not a deterministic gyrochonological relationship applicable to age-dating single stars. From our analysis the rotation rate of one rapidly rotating star with ($G - G_{\rm RP}) \geq 1.25$ is not sufficient to constrain its age, as cluster objects from 10 Myr to 700 Myr and $\sim$6 Gyr field objects at this color are all neighbors in color--period space.

So while traditional gyrochronology relationships rely on evolution after convergence to the slow rotator sequence, an equally valid question for the mid to late M dwarf regime is when such convergence occurs. The oldest clusters with good M dwarf rotation periods at 750 Myr show convergence to the slow rotator sequence at the start of the mid Ms. In field late Ms there are objects that are no longer rapid rotators and that have evolved to longer rotation periods. It is inappropriate to describe it as a single converged slow rotator sequence as the un-associated objects will have a wide range of ages in the field, but this implies that a transition to slow rotation does occur in late Ms. Assigning an age to this transition from rapid to slow rotation would be a valuable tool for dating late Ms, and could lead to a true gyrochronology relation for the population.

The amount of time that late M dwarfs spend rapidly rotating is not known. They must spin down sometime after $\approx$700 Myr, as they have not converged in Praesepe or the Hyades. Sufficiently old coeval populations of stars might reveal an elbow similar to Praesepe at redder colors, but observations of older groupings prove challenging due to their distance and the intrinsic faintness of M dwarfs (known clusters with ages $>$1~Gyr tend to be quite distant, placing their low-mass members out of reach from existing facilities). Without an association of objects of the same age one can turn to field objects with a range of ages. Again, as was shown in Section~\ref{6rotation_age}, the majority of field mid to late M dwarf objects with observed rotation rates are either rapid rotators or have long periods, and so are thought to be on either side of this transition. The bimodality in the field population implies that the transition is quite quick \citep{newton2016}.

Even with understanding when mid and late Ms transition to the slow sequence, understanding the evolution of all M dwarfs after convergence will still be challenging, as current models do not describe the evolution we currently see. \citep{matt_2015} proposes an empirical mass scaling for the torque on a star. However it struggles in the later M's to reproduce the sharp transition from the rapid to slow rotators. \citep{garraffo_2018_revrev} creates a model that uses decreasing magnetic field complexity with rotation rate to explain the bimodality in open clusters. This is a satisfying explanation, as the slow rotators noted in \ref{fig:grid_plot} in each cluster are mostly inacitve in H$\alpha$, which could be a reflection of that decreased magnetic complexity.

It can be seen in the Kepler sample \citep{mcq2014} that there appears to be an upper bound on the rotation rate of more massive F, G, K stars. In \citet{vansaders_2016} this is attributed to a transition of the global stellar dynamo at some point in the stellar lifetime that dramatically reduces the efficiency of magnetic braking. In \citet{vansaders_2016} and \citet{metcalfe_2019} they use a small sample for which rotation rates and other age estimates are available to imply a critical Rossby number at which magnetic activity and stellar rotation decouple, resulting in the rotation spin down essentially stopping. \citet{vansader2019} forward models the Kepler field, and finds the \textit{Rossby edge} at $Ro_{\rm thresh}$ = 2.08. Rotation rates above this are either undetected or absent. They point out that below $T_{eff}$ $\sim$5100\,K rotation periods should be viable even in the oldest stars as the larger convective turn over times will prevent this decoupling. Therefore if a relation can be described and rotation rates remain on the order of 100s of days, M dwarf gyrochronology may be an extremely accurate tool for stellar ages.

%%%%%%%%%%%%%%%%%%%%%%%%%%%%%%%%%%%%%%%%%%%%%%%%%%%%%%%%%%%%%%%%%%%%%%%%%%%%%%%%%%%%%%%%%%%%%%%%%%%%%%%%%%%%%%%%%%%%%%%%
\section{Conclusions}\label{sec:conclusions}

In this work we measured \ksdssrots new M dwarf rotation rates. These objects come from a cross-match of the MLSDSS sample from \citet{kiman_2019} and those observed by the K2 mission. Of them we find evidence for \binary binary systems, based on their light curves, position on the Gaia DR2 CMD, and in some cases the discrepancy between published spectroscopic and parallax distances.

We combine rotation rates from various literature sources and cross match them with Gaia DR2 to create a CMD of \litrots objects. Rotation rate as a function of position on the CMD is dependent on age across most of the main sequence, the one exception being that for $\grp = 1.25-1.4$ there are fast rotators from every sample.

For rotation period distribution across color, we find that in cluster samples rotation periods increase with \grp\ color when objects are on the slow rotator sequence. The youngest clusters have no objects on the slow rotator sequence in the M dwarf regime, while the elbow where objects do join the slow rotator sequence moves redward with age for older clusters, reaching $\grp = 1.25$ for Praesepe/Hyades age of $\approx$700~Myr, with the $\prot \sim 30$~days. Field objects generally have bi-modal rotation distributions, with peaks as rapid rotators $\prot < 2$~days or $>$60 days, with the longest $>$100 days.

For clusters with H$\alpha$ observations, we find that objects with faster rotation rates have larger H$\alpha$ equivalent widths than those with slower rotation rates. Objects in clusters transition from active to inactive in H$\alpha$ at the \grp\ color defined by the elbow of the color--period distribution. Kinematics of the field samples show that velocity dispersion increases with rotation period after $\sim$10 days for the early Ms in the Kepler sample, as well as in the other field samples. There is also evidence of additional kinematic heating at faster rotation rates for the latest objects, which could be due to stalling at fast rotation rates allowing for additional heating before transitioning to longer rotation periods.

To accurately date the transition of these objects across age and color will likely require statistical modeling of field objects that incorporate other age dating techniques. Examples of these techniques include using H$\alpha$ activity and kinematic velocities, which have both been discussed in this work in section \ref{7halphakin}. If a large enough sample distributed across color is developed, comparison of activity levels or kinematic properties of objects on either side of the transition could constrain the age at which the evolution occurs. Wide binaries where the M dwarf's age can be determined by the companion (be it a more massive main-sequence star, an evolved star, or a white dwarf), can also be used to tighten these constraints further. Calibrating the elbow with color through the mid and into the late M dwarfs is a likely first step for establishing the gyrochronology for the M dwarf regime.

%%%%%%%%%%%%%%%%%%%%%%%%%%%%%%%%%%%%%%%%%%%%%%%%%%%%%%%%%%%%%%%%%%%%%%%%%%%%%%%%%%%%%%%%%%%%%%%%%%%%%%
\acknowledgments

Support for this work was provided by the William E Macaulay Honors College of The City University of New York.
This work has been supported by NASA K2 Guest Observer program under award 80NSSC19K0106. 
This material is based upon work supported by the National Science Foundation under Grant No. 1614527. JKF acknowledges support from the Heising Simons foundation and the Research Corporation for Science Advancement (Award 2019-1488).

This paper includes data collected by the Kepler and 
K2 missions, 
which are funded by the NASA Science Mission directorate.
We obtained these data from the Mikulski Archive for Space Telescopes (MAST). 
STScI is operated by the Association of Universities for Research in Astronomy, Inc., 
under NASA contract NAS5-26555. 
Support for MAST for non-HST data is provided by the NASA Office of Space Science via 
grant NNX09AF08G and by other grants and contracts.

This work has made use of data from the European Space Agency (ESA)
mission {\it Gaia} (\url{https://www.cosmos.esa.int/gaia}), processed by
the {\it Gaia} Data Processing and Analysis Consortium (DPAC,
\url{https://www.cosmos.esa.int/web/gaia/dpac/consortium}). Funding
for the DPAC has been provided by national institutions, in particular
the institutions participating in the {\it Gaia} Multilateral Agreement.

This work made use of the \url{gaia-kepler.fun} cross-match database created by Megan Bedell.

This research has also made use of NASA's Astrophysics Data System, 
and the VizieR \citep{vizier} and SIMBAD \citep{simbad} databases, 
operated at CDS, Strasbourg, France.

\facilities{Gaia, Kepler, K2}

\software{astropy \citep{astropy_2013,astropy_2018}, exoplanet \citep{exoplanet}, EVEREST \citep{Luger2018}}

\bibliography{main}

\end{document}